\documentstyle[12pt]{article}

\begin{document}
\renewcommand{\thefootnote}{\fnsymbol{footnote}}


\begin{center}

{\large \bf SPINS AND CHARGES, THE ALGEBRA AND SUBALGEBRAS OF
THE GROUP SO(1,14) AND GRASSMANN SPACE } 

 
\vspace{5mm}

 NORMA MANKO\v C BOR\v STNIK  and SVJETLANA FAJFER 

\vspace{3mm}

{\it Department of Physics, University of Ljubljana, Jadranska 19, }\\
{\it J. Stefan Institute, Jamova 39, 61 111 Ljubljana, Slovenia }\\

\vspace{12mm}

\newpage

{\large ABSTRACT }

\end{center}

\vspace{3mm}

In a space of $d=15 $ Grassmann coordinates, two types of
generators of the Lorentz transformations, one of spinorial and
the other of vectorial character, both linear operators in 
Grassmann space, forming the group $
SO(1,14) $ which contains as subgroups $ SO(1,4) $ and $ SO(10)
$ ${\supset SU(3)} { \times SU(2)} { \times U(1)} $, define
the fundamental and the adjoint representations of the group,
respectively. The eigenvalues of the commuting operators can be
identified with  spins of fermionic and bosonic fields $
(SO(1,4)) $, as well as with their Yang-Mills charges $ (SU(3)$,
$ SU(2)$, $ U(1)) $, offering  the unification  of not only all Yang
- Mills charges but  of all
the internal degrees of freedom of fermionic and bosonic fields
-  Yang - Mills charges and spins - and accordingly of all
interactions - gauge fields and gravity. The theory suggests that
elementary particles are either in the "spinorial"
representations with respect to spins and all 
charges, or they are in the "vectorial" representations with
respect to spins and all charges, which indeed is the case with
the quarks, the leptons and the gauge bosons. 

The algebras of the two
kinds of generators of  Lorentz transformations in  Grassmann
space were studied and the representations are commented on.

\vspace{3mm}
\vspace{1cm}

\newpage


{\bf 1. Introduction. }    

\vspace{0.3cm}

The fact that  ordinary space-time is not enough to
describe  dynamics of particles, was  recognized first in 1925
 when in addition 
to the vector space, spanned over the ordinary coordinate space,
a space of two vectors, called the internal space of the
fermionic spin,  
was introduced in order to 
describe the 
spin ($\frac{1}{2}$) of fermions \cite{uhl}. In 1928 this
internal space was enlarged to four vectors, including the
particle-antiparticle degrees of freedom in order to describe
relativistic fermions  \cite{dirac}. 
In 1932  to  the internal space of  fermionic spins the space of 
two vectors describing the fermionic
isospin, or how it is  also  called now - the weak charge of
fermions, was added 
\cite{heis}, and in 1964 to these two spaces the
internal space of three vectors describing the colour charge of
fermions was added\cite{gel}
. 
The unification of electromagnetic and weak interactions makes
clear that also the 
electromagnetic charge origins in the internal 
space \cite{wein,salam,glash}. 

All symmetries in physics are in  theories  connected with
the appropriate groups. Spins  are connected with 
the Lorentz group $ SO(1,3) $ ( in the three
dimensional subspace of the four dimensional space time spins
are described by the group $ SU(2) $ ), while charges are
connected with the group $U(1)$ (
the electromagnetic charge ), the group 
$ SU(2) $ ( the weak charge ) and the group $ SU(3) $ ( the
colour charge ).

While in ordinary space time the realization of vectorial types
of representations for the Lorentz group ( intiger angular
momenta ) only are possible, for the internal spaces two types
of representations are 
required: the fundamental and the adjoint. The fundamental
representations are used to describe the above mentioned
properties of fermions, the adjoint 
representations are used to describe the corresponding
properties of bosons - the gauge vector fields. 

To both types of
representations two types of singlets
have to be added.

Modern theories: the Standard Electroweak Model \cite
{glash,salam,wein}, the Grand
Unification Theories \cite{georg}, the String Theories
\cite{kaku}, the Kaluza-Klein
Theories \cite{duff}, the Technicolour Models \cite{susk}, the
Supersymmetric \cite{wess} Theories 
try to unify the  internal spaces of charges, but
they do not unify the internal spaces of charges  with the
internal space of spins.

This paper elaborates the idea \cite{man1,man2} of {\it 
unifying all the internal degrees of freedom - spins and
charges.}

The space of $d$ ordinary commuting and $d$ Grassmann
anticommuting coordinates, $ d \ge 15 $, offers the possibility
of describing all degrees of freedom
of particles which today are 
supposed to be elementary.
All internal degrees of freedom are in this space described as
the dynamics 
in  the Grassmann part of  space. 

{\it Two
kinds of generators of the Lorentz transformations can be
defined in Grassmann  space: one of spinorial character }( spinorial
operators,) 
{\it the other of vectorial character} ( vectorial operators.)  For $
d=15$, contain the representations, if determined by 
the  operators forming the subgroups $
SO(1,3), SU(3), SU(2), U(1) $ of the group $ SO(1, 14)$,  all 
the vectors needed to describe  quarks, leptons and
all gauge fields. 
Spinorial representations, including fundamental representatins and
singlets, describe spins and charges of 
fermions, while vectorial representations, including adjoint
representations and singlets, describe spins
and charges of bosons.

{\it Spins and charges are in the proposed theory unified}. As the
consequence, in this theory {\it elementary particles should
either be in the 
representations, defined by the spinorial operators of 
the Lorentz group $ SO(1,3) $ and all the groups describing
charges, or they should be in the representations, defined by
the vectorial operators of the Lorentz group and all the
groups describing charges}.

This is in agreement with the properties of known
quarks, leptons and  gauge bosons since
they are either fermions in the 
fundamental representations with respect to the Lorentz group
and the groups describing charges or they are singlets with
respect to these groups, or they
are bosons in the adjoint representations with respect to the
Lorentz group and the groups describing charges or they are
singlets with respect to these groups.

The Higgs's boson of the Standard Electroweak Model can in
the proposed approach be described as a weak doublet or a 
constituent field, made 
of two fermions,  one  singlet and one doublet with
respect to the group $ SU(2) $ describing the weak charge.

The supersymmetric partners of  
weak bosons and  gluons, for example, suggested
by the minimal supersymmetric extension of the Standard
Electroweak Model \cite{raby} to be fermions in the adjoint
representations with 
respect to the groups describing charges, can appear in the
proposed theory as 
constituent particles only.

The purpose of this article is to show that generators 
of the Lorentz transformations in  $ d=2n+1$  
dimensional Grassmann space, forming the Lie algebra of the
Lorentz group $ SO(1,2n) $ with the  subalgebra 
$ SO(2n-4) {\times SO(1,4)}$, define for the choice n=7 the
Lorentz subalgebras $SO(1,4)$ and  $SO(10)$. Generators of
$SO(1,4)$, if having spinorial character, determine
spinors or, if having vectorial character, determine scalars and
vectors in the four dimensional subspace. Generators of $SO(10)$
subalgebra, being decomposed into generators of $SU(3), SU(2) $
and $ U(1) $, accordingly determine colour, weak and 
electromagnetic charges for spinors or for scalars and vectors. 

We study the
decompositions of  generators of  subgroups in terms of 
generators of the Lorentz transformations of spinorial
and vectorial character.

All the generators are differential
operators in Grassmann space. We look for  Casimir operators of 
subgroups. Solving the eigenvalue problem for commuting
operators, we look for 
representations of groups and subgroups, defined by both 
kinds of  generators of the Lorentz transformations in 
Grassmann space, separately, and comment on them. 

The theory offers a way of looking for algebras, subalgebras and
their representations. It contains a parallel method to Dynkin
diagrams and Young tableaux.

\vspace{4mm}

{\bf 2. Coordinate Grassmann space and linear operators. } 

\vspace{0.3cm}

In this section we briefly repeat a few definitions concerning
a d-dimensional Grassmann space,  linear Grassmann space 
spanned over the coordinate space, linear operators defined in
this space and the Lie algebra of generators of the Lorentz
transformations \cite{berez,lurie} . 

\vspace{3mm}

{\it 2.1. Coordinate space with Grassmann character}

\vspace{3mm}

We define a d-dimensional Grassmann space of real anticommuting
coordinates $ \{ \theta ^a \} $, $ \theta^{a*} = \theta^a $, $
a=0,1,2,3,5,6,...,d,$ 

satisfying the anticommutation relations

$$ \theta^a \theta^b + \theta^b \theta^a := \{ \theta^a, \theta^b
\} = 0,
\eqno (2.1) $$

called the Grassmann algebra \cite{man2,berez}. The metric tensor $ \eta
_{ab}$ $ = diag (1,$ $ -1, -1, -1,..., -1) $ lowers the indices of a
vector $\{ \theta^a \} = \{ \theta^0, \theta^1,..., \theta^d \},
\theta_a = \eta_{ab} \theta^b$. Linear transformation actions on
vectors $ (\alpha \theta^a + \beta  x^a) $

$$ (\alpha \acute{\theta}^a + \beta \acute{x}^a ) = L^a{ }_b
(\alpha \theta^b + \beta x^b ) ,\eqno (2.2) $$

which leave forms

$$ ( \alpha \theta^a + \beta x^a ) ( \alpha \theta ^b + \beta
x^b ) \eta_{ab} \eqno (2.3) $$

invariant, are called the Lorentz transformations. Here $
(\alpha \theta^a + \beta x^a ) $ is a  vector of d
anticommuting components (Eq.(2.1)) and d commuting $ (x^ax^b -
x^bx^a = 0) $ components, and $ \alpha $ and $ \beta$ are two
complex numbers. The requirement that forms (2.3) are scalars with
respect to the linear transformations (2.2), leads to the
equations 

$$ L^a{ }_c L^b{ }_d \eta_{ab} = \eta_{cd}.  \eqno(2.4)$$

{ 2.2 \it Linear vector space.}

A linear space spanned over a Grassmann coordinate space of d
coordinates has the dimension $ 2^d$. If monomials $
\theta^{\alpha_1} \theta^{\alpha_2}....\theta^{\alpha_m} $
are taken as a set of basic vectors with $\alpha_m \neq
\alpha_j,$ half of the vectors have an even (those with an even m)
and half of the vectors have an odd (those with an odd n)
Grassmann character. Any vector in this space may be represented
as a linear superposition of monomials 

$$ f(\theta) = \alpha_0 + \sum_{i=1}^{d}  \alpha _{a_1a_2 ..a_i}
\theta^{a_1} \theta^{a_2}....\theta^{a_i},\;\; a_k< a_{k+1},
\eqno(2.5)$$ 

where constants $\alpha_0, \alpha_{a_1a_2..a_i}$ are complex
numbers. 

\vspace{3mm}

{ 2.3 \it Linear operators. }

\vspace{3mm}
 
In  Grassmann space the left derivatives have to be
distinguished  from the right derivatives, due to the 
anticommuting nature of the coordinates \cite{man1,man2,berez}.
We shall make use 
of left derivatives $ {\overrightarrow {{\partial}^{\theta}}}{
}_a:= \overrightarrow {\frac {\partial}{\partial
\theta^a}},\;\;\; {\overrightarrow {{\partial}^{\theta}}}{ }^{a}:=
{\eta ^{ab}} \overrightarrow {{\partial}^{\theta}}{ }_b \; $, on
vectors of the 
linear space of monomials $ f(\theta)$,  defined as follows:  

$$ {\overrightarrow{{\partial}^{\theta}}}{ }_a\; \theta^b f(\theta):
= \delta^b{ }_a f(\theta) - \theta^b  
{\overrightarrow{{\partial}^{\theta}}}{ }_a\; f(\theta) , \eqno(2.6) $$

$$ {\overrightarrow {{\partial}^{\theta}}}{ }_a \alpha f(\theta):=
(-1)^{n_{a \partial}} \alpha {\overrightarrow{{\partial}^{\theta}}}{
}_a \;f(\theta).$$

Here  $ \alpha $ is a constant of either commuting $( \alpha
\theta^a - \theta^a \alpha = 0 )$ or anticommuting $( \alpha
\theta^a + \theta^a \alpha = 0 )$ character, and 

$$ n_{AB} = \left\{ \begin{array} {ll} +1, & {\rm if\; A\; and\; B \;
have\;Grassmann\; odd\; character}\\
0, &{\rm otherwise.} \end{array} \right\} $$

We define the following linear operators \cite{man1,man2}

$$ p^{\theta} { }_a := -i {\overrightarrow{{\partial}^{\theta}}}{
}_a , \;\;
 \tilde{a} ^a := i(p^{\theta a} - i \theta^a) ,\;\;
\tilde{\tilde{a}}{}^a := -(p^{\theta a} + i \theta^a). \eqno
(2.7) $$

If the inner product is defined as in Sect. (4.1.)  it follows 

$$ \theta^{a}+ = i \eta^{aa} p^{\theta a}, p^{\theta a}+ = i
\eta^{aa} \theta^a, \tilde a^{a}{ }^+ = \eta^{aa} \tilde
a^a,\;\;\; \tilde{\tilde 
a}{ }^{a}{ }^+ =  \eta^{aa} \tilde{\tilde a}{ }^{a}. \eqno(2.7a)
$$

We define the generalized commutation relations \cite{man1,man2}:

$$ \{ A,B \} := AB - (-1) ^{n_{AB}} BA, \eqno(2.8) $$

fulfilling the equations

$$ \{A,B\}  = (-1)^{n_{AB}+1} \{B,A\},                  \eqno(2.9a)$$  

$$ \{A,BC \} = \{A,B\} C + (-1)^{n_{AB}}B \{A,C\},      \eqno(2.9b)$$ 

$$ \{AB,C\} = A\{B,C\} + (-1)^{n_{BC}}\{A,C\}B,
                                     \eqno(2.9c)$$

$$ (-1)^{n_{AC}}\{A,\{B,C\}\} + (-1)^{n_{CB}}\{C,\{A,B\}\}
+ (-1)^{n_{BA}} \{B,\{C,A\}\} = 0.                       \eqno(2.9d)$$

We find

$$ \{p^{\theta a}, p^{\theta b} \} = 0 = \{ \theta^{a},
\theta^{b}\}, \eqno(2.10a)$$ 

$$  \{p^{\theta a}, \theta^{b}\} = -i \eta^{ab},         \eqno(2.10b)$$
$$ \{\tilde{a}^{a}, \tilde{a}^{b} \} = 2 \eta^{ab}
= \{\tilde{\tilde{a}}{ }^{a}, \tilde{\tilde{a}}{ }^{b} \},      \eqno(2.10c)$$

$$ \{ \tilde{a}^{a}, \tilde{\tilde{a}}{ }^{b} \} = 0.       \eqno(2.10d)$$

We see that $\theta ^a $ and $ p^{\theta a} $ form a Grassmann
odd Heisenberg algebra, while $ \tilde a^a $ and $
\tilde{\tilde{a}}{ }^a $ form the Clifford algebra.

\vspace{4mm}

{ 3. \bf Lie algebra of generators of Lorentz transformations. }

\vspace{0.3cm}

We define two kinds of operators \cite{man1,man2}. The first ones are
binomials of operators forming the Grassmann odd Heisenberg
algebra 

$$ S^{ab} : = ( \theta^a p^{\theta
b} - \theta ^b p^{\theta a} ).                     \eqno  (3.1a)$$

The second kind are binomials of operators forming the Clifford
algebra 

$$ \tilde S ^{ab}: = - \frac{i}{4} [\tilde a ^a , \tilde a ^ b
], \;\; \tilde {\tilde S} { }^{ab}: = - \frac{i}{4} [ \tilde
{\tilde a}{ }^a , \tilde {\tilde a}{ }^b ] , \eqno(3.1b)$$ 

with $ [A, B]:= AB - BA.$

Either $ S^{ab} $ or $ \tilde S ^{ab}$  or $ \tilde{\tilde S}{
}^{ab} $ fulfil the Lie algebra of the Lorentz group $ SO(1,d-1)
$ in the d-dimensional Grassmann space :

$$ \{ M^{ab}, M^{cd} \} = -i ( M^{ad} \eta^{bc} + M^{bc}
\eta^{ad} - M^{ac} \eta^{bd} - M^{bd} \eta^{ac} ) \eqno(3.2)$$

with $ M^{ab} $ equal either to $ S ^{ab} $ or to $\tilde S
^{ab} $ or to $ \tilde {\tilde S} { }^{ab} $ and $ M^{ab} = -
M^{ba} $. There are $ d (d-1)/2 $ operators of each kind in  d
- dimensional Grassmann space.  

We see that 

$$ S^{ab} = \tilde S ^{ab} + \tilde {\tilde S}{ }^{ab},\;\;
 \{ \tilde S ^{ab} , \tilde {
\tilde S }{ }^{cd} \} = 0 = \{ \tilde S ^{ab} , \tilde {\tilde
a}{ }^c \} = \{ \tilde a ^a , \tilde { \tilde S }{ }^{bc} \}.
\eqno (3.3)$$ 

The operators $ \tilde S ^{ab} $, as well as
the  operators $ \tilde {\tilde S}{ }^{ab} $,  define
what we call  the
spinorial representations of the Lorentz group $ SO(1,d-1) $
and of  subgroups of this Lorentz group, 
 while $ 
S^{ab} = \tilde S ^{ab} + \tilde{\tilde S} { }^{ab} $ 
 define what we call  the vectorial representations of
the Lorentz group $ SO(1,d-1)$
and of subgroups of this group.\cite{man1,man2}
The spinorial representations contain what is called the
fundamental representations of this Lorentz group and of
subgroups and they contain also singlets, while the vectorial
representations contain what is 
called the regular or the adjoint representations 
of this Lorentz group and of subgroups and they contain also
singlets. 
Group elements are in any of the three cases defined by:
 
$$ {\cal U}(\omega) = e^{ \frac{i}{2} \omega_{ab} M^{ab}} 
,\eqno(3.4) $$

where $ \omega_{ab} $ are the parameters of the group.

Linear transformations, defined in Eq.(2.2), can then be written
in terms of group elements as follows

$$ \acute{\theta}^a = L^a{ }_b \theta ^b = e^{- \frac{i}{2}
\omega_{cd} S^{cd}} \theta ^a e^{\frac{i}{2} \omega _{cd}
S^{cd}}, \eqno(3.5a) $$

$$ \acute{\tilde a}^a = L^a{ }_b \tilde a ^b = e^{- \frac{i}{2}
\omega_{cd} S^{cd}} \tilde a ^a e^{\frac{i}{2} \omega _{cd}
S^{cd}}, \eqno(3.5b) $$ 

$$ \acute{\tilde{\tilde a}}{ }^a = L^a{ }_b \tilde{\tilde a}{ }
^b = e^{- \frac{i}{2} 
\omega_{cd} S^{cd}} \tilde{\tilde a}{ } ^a e^{\frac{i}{2} \omega _{cd}
S^{cd}}, \eqno(3.5c) $$

where  $\omega _{cd} $ are  the parameters of the transformations.
Since $ \{ \tilde S^{ab} , \tilde{\tilde S}{ }^{cd} \} = 0 $, 
in Eq.(3.5b) $ S^{cd} $ can be replaced by $ \tilde S ^{cd} $
and in Eq.(3.5c) $ S^{cd} $ can be replaced by $ \tilde {\tilde
S }{ }^{cd} $.

 By using Eqs.(2.9) and (3.2) it can be proved for any d , that
$ M^2 $ is the invariant of the Lorentz group 

$$ \{ M^2, M^{cd} \} = 0,\;\; M^2 =  \frac{1}{2} M^{ab} M_{ab} ,
 \eqno(3.6) $$ 

and that for d=2n we can find the additional invariant $ \Gamma $

$$ \{ \Gamma, M^{cd} \} = 0,\;\; \Gamma = \frac{i(-2i)^{n}
}{(2n)!} 
\epsilon_{a_1a_2...a_{2n}} M^{a_1a_2} ....M^{a_{2n-1}a_{2n}} , 
 \eqno(3.7) $$

where $\epsilon _{a_1a_2...a_{2n}} $ is the totally antisymmetric
tensor with $ 2n $ indices and with $ \epsilon _{ 1 2 3 ...2n }
\;\; = 1 $. This means that $ M^2  $ 
and $ \Gamma $ are for $ d = 2n$ the two invariants or Casimir
operators of the group $ SO(d) $ (or $ SO(1,d-1) $, the
two algebras differ 
in the definition of the metric $ \eta^{ab} $).
For  $ d = 2n + 1 $ the 
second invariant cannot be defined.
( However, for $ d = 2n + 1 $ one can still define the invariants $
\tilde \Gamma = 
\prod_{a=1,..,d} \sqrt{\eta^{aa}} \tilde a{ }^a $ and $ \tilde
{\tilde \Gamma} = 
\prod_{a=1,..,d} \sqrt{\eta^{aa}} \tilde{\tilde a}{ }^a $, which
commute with $ \tilde S^{ab} $ and $\tilde{\tilde S}{ }^{ab} $.)

While the invariant $ M^2 $ is trivial in the
case when $ M^{ab} $ has spinorial character, since  
 $ (\tilde S^{ab})^2 = \frac{1}{4} \eta^{aa} \eta^{bb} = 
(\tilde {\tilde S}{ }^{ab})^2 $ and therefore $ M^2 $ is equal
in both cases  to the number $ \frac{1}{2} \tilde
S^{ab} \tilde S _{ab} = \frac{1}{2} \tilde {\tilde S}{ }^{ab}
\tilde {\tilde 
S}{ }_{ab} = d ( d-1 ) \frac{1}{8}$ , it is a nontrivial
differential operator in  Grassmann space if $ M^{ab}$
have  vectorial character $(M^{ab} = S^{ab})$. The invariant of
Eq.(3.7) is always a nontrivial operator.

We shall discuss the representations of groups in Sect.4.

\vspace{0.3cm}

{ 3.1. \it Algebras of subgroups of the Lorentz groups }

\vspace{3mm}

In this section we shall present some subalgebras of the
algebras of  the groups $
SO(d) $ or $ SO(1,d-1) $ for a few chosen $ d $. We shall choose
subalgebras  appropriate for the description of the
internal degrees of freedom of fermionic and bosonic
fields, that is for spins and known Yang-Mills charges.

We shall present operators forming the desired subalgebras in
terms of generators $ M^{ab} $, which have either the spinorial
character - $ \tilde S^{ab} , \tilde{\tilde S}{ }^{ab} $ - and
define  spinorial representations, or  they have the 
vectorial character - $ S^{ab} $ - and define 
 vectorial representations of the group. We shall analyse
subalgebras of $ SO(d) $ or $ SO(1,d-1) $ for $ d = 4,5,6,8,15
$. In the case of
the group $ SO(d) $ the matrix $ \eta^{ab} $ has all the
diagonal elements equal to -1.

It is selfevident that {\it the algebra} $ SO(1,d-1) $ {\it or} $
SO(d) $ {\it 
contains} \cite{man1,man2}  $ N $ {\it subalgebras defined by 
operators} $ \tau ^{\alpha i}, \alpha = 1,N, \;\; i = 1,N_{\alpha} $,
{\it where} $ N_{\alpha } $ {\ is the number of elements of each
subalgebra, with the properties }

$$ [ \tau ^{\alpha i} , \tau ^{\beta j} ] = i \delta ^{\alpha
\beta} f^{\alpha ijk } \tau ^{\alpha k}, \eqno (3.8) $$

{\it if operators} $ \tau ^{\alpha i} $ {\it can be expressed as
linear superpositions  of operators} $ M^{ab} $ 

$$ \tau^{\alpha i} = {\it c} ^{\alpha i} { }_{ab} M^{ab}, \;\;
{\it c} ^{\alpha i}{ }_{ab} = - {\it c} ^{\alpha i}{ }_{ba}, \;\;
\alpha=1,N, \;\;
i=1,N_{\alpha}, \;\;a,b=1,d. \eqno(3.8a) $$

Here $ f^{\alpha ijk} $ are structure constants
of the ($ \alpha $) subgroup with $ N_{\alpha} $ operators.
According to the three kinds of operators $ M^{ab} $, two of
spinorial and one of vectorial character, there are  three kinds
of operators $ \tau^{\alpha i} $ defining subalgebras of
spinorial and vectorial character, respectively. All three kinds
of operators are, according to Eq.(3.8), 
defined by the same coefficients $ {\it c}^{\alpha i} { }_{ab} $
and the same structure constants $ f^{\alpha i j k } $.
From Eq.(3.8) the following relations among constants ${\it
c}^{\alpha i}{ }_{ab} $ follow:

$$ -4 {\it c}^{\alpha i}{ }_{ab} {\it c}^{\beta j b}{ }_c -
\delta^{\alpha \beta} f^{\alpha ijk} {\it c}^{\alpha k}{ }_{ac}
= 0. \eqno (3.8b)$$

In the case when the algebra and the chosen subalgebras are
isomorphic, that is if the number of  generators of
subalgebras is equal to $ \frac{d(d-1)}{2} $ , the inverse
matrix $ e^{\alpha iab} $ to the matrix of coefficients $
c^{\alpha i}{ }_{ab} $ exists \cite{man2}

$$M^{ab} = \sum_{\alpha i} e^{\alpha iab} \tau^{\alpha i},
\eqno(3.8c) $$

with the properties $ c^{\alpha i}{
}_{ab} e^{\beta jab} = \delta^{\alpha \beta} \delta^{ij}, \;\;
c^{\alpha i}{ }_{cd} e^{\alpha iab} = \delta^a{ }_c \delta^b{
}_d - \delta^b{ }_c \delta^a{ }_d $.

When we look for coefficients $ c^{\alpha i}{ }_{ab} $ which
express operators $ \tau ^{\alpha i} $, forming a subalgebra
$ SU(n) $ of an algebra $ SO(2n) $ in terms of $ M^{ab} $, the
procedure is rather simple. For  spinorial representations
we define Grassmann odd operators $\tilde b^i $ and their
hermitian conjugate ( Eq.(2.7b)) $ \tilde b^{i+} $

$$ \tilde b ^i = \frac{1}{2} ( \tilde a^{(2i-1)} - i \tilde
a^{(2i)} ) ,\;\;\; \tilde b ^{i}{ }^+ = \frac{1}{2} ( \tilde a^{
(2i-1)}{ }^+ + i \tilde a^{ (2i)}{ }^+ ). \eqno(3.9a)  $$ 

We take the traceless matrices $ (\tilde \sigma{ }^{\alpha
m})_{ij} $ which form the algebra of $ SU(n) $ :$ (\tilde \sigma
{ }^{\alpha m} )_{ij} (\tilde \sigma
{ }^{\alpha n} )_{jk} - (\tilde \sigma
{ }^{\alpha n} )_{ij}(\tilde \sigma
{ }^{\alpha m} )_{jk} = i f^{\alpha mnl} (\tilde \sigma{ }^{\alpha
l})_{ik}.$ Here $ f^{\alpha mnl} $ are structure constants of
the group $ SU(n)$.  We then
construct  operators $\tilde \tau^{\alpha m}$ as follows

$$ \tilde \tau^{\alpha m} = (\tilde b^j)^+ ( \tilde
\sigma^{\alpha m} )_{jk} \tilde b^k =$$

$$ \frac{1}{4} ( \tilde \sigma^{\alpha m} )_{jk} ( 2 \eta_{jk} + \tilde
a^{+(2j-1)} \tilde a^{(2k-1)} + \tilde
a^{+(2j)} \tilde a^{(2k)} + i \tilde
a^{+(2j)} \tilde a^{(2k-1)} - i\tilde
a^{+(2j-1)} \tilde a^{(2k)} ),$$

$$ \;\; \alpha = \{1,N \} \;\;\;\; m = \{1,n^2
-1\},\;\; \{j,k\} = \{1,n\}. \eqno(3.9b) $$ 
 
Since  $ \tilde a^{a}{ }^+ = \eta^{aa} \tilde a^a $ (See Eq.(2.7b)) and
$(\tilde \sigma^{\alpha m} )_{jk} $ are traceless matrices, we
find for the group $SO(2n)$: 

$$ \tau^{\alpha m} = -\frac{i}{2} (\tilde \sigma^{\alpha m})_{jk}
 \{ M^{(2j-1) (2k-1)} +
M^{(2j) (2k)} + i M^{(2j) (2k-1)} - i M^{(2j-1) (2k)}
\}.\eqno(3.9c)$$

One can easily prove, if taking into
account Eq.(3.2), that operators $ \tau^{\alpha m} $ fulfil the
algebra of the group $ SU(n) $ (Eq.(3.8)), for any of three
choices for operators $ M^{ab} : S^{ab}, \tilde S^{ab},
\tilde{\tilde S}{ }^{ab}$. We have generalized expressions for
operators $ \tilde \tau^{\alpha m} $, which have spinorial
character, to all three kinds of operators. Eq. (3.9c) can
be found in ref. \cite {georg2}. ( Instead of Grassmann odd
operators $\tilde 
b^a$ which are written in
terms of Grassmann odd operators $\tilde a^a $, we
could define Grassmann even operators in terms of $\tilde
\gamma^a$ operators which we shall introduce in the next subsection.
Then in a similar way as for spinorial operators $\tilde
\tau ^{\alpha m} $, the procedure to find $\tau ^{\alpha m}$
for vectorial operators can be
found.)

In this section  we shall present coefficients ${\it c}^{\alpha i}{
}_{ab} $ for 
a few cases which seem to be interesting for  particle
physics. As have we already  said, the coefficients are the same
for all three kinds of operators, two of spinorial and one of
vectorial character.  The representations, of course, depend on the
operators.

We shall use  the
notation $\tilde \tau^{\alpha i} $ or $\tilde{\tilde \tau}{
}^{\alpha i} $ for  operators
$\tau^{\alpha i} $ (Eq.(3.8a)) when they have  spinorial
character (in such a case  operators $M^{ab}$
have to be replaced by the corresponding operators of spinorial
character $\tilde S^{ab}$ or $\tilde{\tilde S}{ }^{ab}$ ) . We
shall use, however, the same notation as in 
the general case 
$\tau^{\alpha i}$, if  operators $ M^{ab} $ describe vectorial 
character and have therefore to be expressed in terms of $ S^{ab} $.

\vspace{3mm}

{ 3.1.1. \it Subgroups of SO(1,3) }

\vspace{3mm}

This problem is discussed for spinorial degrees
of freedom in many text-books \cite {georg2,lurie,ham,grein}. 
We shall follow here 
ref.\cite{man2} where 
the   algebra of $ SO(1,3) $ is presented together with the 
corresponding irreducibile representations for operators of
spinorial  and vectorial character in  Grassmann space. 

There are six generators $ ( d(d-1)/2 ) $ of the group $
SO(1,3) $  each of  
three kinds ($ M^{ab}$ is either $ S^{ab} $ or $ \tilde{S}^{ab}
$ or $ \tilde{\tilde S}{ }^{ab} $ ). We find according to Eqs.(
3.6, 3.7) two invariants of the group: $ M^2:= \frac{1}{2} M^{ab}
M_{ab}\; $ and $ \Gamma:= \frac{-i}{3!} \epsilon_{abcd} M^{ab}
M^{cd} $. For $ M^{ab} = \tilde S^{ab} $ or $ M^{ab} =
\tilde{\tilde S}{ 
}^{ab} $ the first invariant is the number $\frac{6}{4}$ ( $
(\tilde S^{ab})^2 = \frac{1}{4} \eta^{aa} \eta^{bb} =
(\tilde{\tilde S}{ }^{ab})^2 $), while  for $ M^{ab} =
S^{ab} $ it is a nontrivial
differential operator in  Grassmann space. The second
invariant is the nontrivial operator in any 
of the three cases.  

If  $ M^{ab}$ is equal to either $ \tilde S
^{ab} $ or to $ \tilde{\tilde S}{ }^{ab} $, the second invariant
$ \Gamma $ can be recognized as  the
operator of chirality, the product of the Dirac $\gamma ^a $
matrices \cite{man2,grein}. To see this one
should write the operator $ {\tilde {\Gamma }}$ in terms of
$\tilde a^a $ or $ \tilde{\tilde a}{ }^a$. One finds the product
of all operators  $ \tilde a^a $: $ {\tilde
\Gamma} = i \tilde a^0 \tilde a^1 \tilde a^2 \tilde a^3 $ or
equivalently $ \tilde{ \tilde {\Gamma}} = i \tilde {\tilde a}{
}^0 \tilde{ \tilde a}{ }^1 \tilde {\tilde a}{ }^2 \tilde{\tilde
a} { }^3 $. 

Operators $ \tilde a^a $ and $\tilde{ \tilde a}{
}^a $ are Grassmann odd operators. Operating on spinors they
would change spinors to vectors or scalars, changing their
Grassmann character from odd to even \cite{man1,man2}. They
cannot, therefore, be recognized as the 
 Dirac $ \gamma ^a $ matrices. In the next subsection we
shall find it meaningful to 
define $ - 2i \tilde S ^{5a} = - \tilde a^5 \tilde a^a$ $
( - 2i \tilde {\tilde 
S}{ }^{5a} = - \tilde {\tilde a}{ }^5 \tilde{\tilde a}{ }^a )\; $
as operators 
corresponding to the Dirac $\gamma ^a $ matrices. Since
according to Eq.(2.10c)  
$ (\tilde a^5)^2 = -1 = (\tilde {\tilde a}{ }^5)^2 $, the
product of all $ \tilde a^a $ is equal to the product of
the corresponding $\tilde \gamma^a $ defining the same
invariant $ {\tilde \Gamma} $ and equivalently for $\tilde
{\tilde \Gamma} $.

The corresponding representations will be discussed
in Sect.4.

It is easy to find $ SO(3) $ as a subgroup of the group $
SO(1,3) $ with three generators $ M_{12},$ $ M_{13},$ $ M_{23} $
closing the Lie subalgebra with the properties 

$$ \tau^{1i} = \frac{1}{2} \epsilon ^{ijk} M^{jk} ,\;\; i,j,k
\in \{ 1,2,3 \}, \eqno (3.10) $$

where  $ f^{1ijk} = \epsilon ^{ijk} $ . In this case we
see, by comparing Eqs.(3.8a) and (3.10) , that

$$ {\it c} ^{1i}{ }_{ab} = {\frac{1}{2} \epsilon ^i{ }_{ab}, \; if
\:a \: and \: b \in \{ 1,2,3 \} \choose 0,\; otherwise }.
\eqno(3.10a) $$ 

The only invariant of the subgroup is $ \frac{1}{2} M^{ij}
M_{ij} $, which is equal to $ \frac{3}{4} $ in the case that $
M^{ij} = \tilde S^{ij} ( \tilde {\tilde S}{ }^{ij}) $ and  is a
nontrivial operator if $ M^{ij} = S^{ij} $. The two Casimir
operators of the group $ SO(1,3) $ are  of course the Casimir
operators of the subgroup $ SO(3) $ as well.

Requiring that  two kinds of operators $\tau^{\alpha
i} $ exist $,\;\; \alpha = 1,2; \;\;
i =1,2,\;\; $ with the structure constants $ f^{\alpha ijk} =
\epsilon ^{ijk} ; \;\; \alpha =1,2 ; \;\; i,j,k ={1,2,3} $, one
easily finds that

$$ {\it c} ^{\alpha i}{ }_{ab} = \left( \begin{array}{cc}
\frac{1}{4} \epsilon ^i {
}_{ab}, & if \; a \; and \; b \in \;\; \{ 1,2,3 \} \\
(-1)^{\alpha+1} \frac{i}{4}, &  if\; a=0 \\
(-1)^{\alpha} \frac{i}{4}, & if\;\; b=0 \end{array} \right).
\eqno(3.10b) $$

Since the two subalgeras are isomorphic to the algebra of $
SO(1,3) $, the matrix $ {\it c}^{\alpha i}{ }_{ab} $ has the
inverse matrix ${\it e}^{\alpha iab} $ (eq.(3.8c))

$$ {\it e} ^{\alpha iab} = \left( \begin{array}{cc}
\epsilon ^{iab}, & if \; a\; and \; b
\in \;\; \{1,2,3 \} \\ 
 i (-1)^{\alpha}, &  if\;\;  a=0 \\ 
i (-1)^{\alpha +1}, & if \;\; b=0 \end{array} \right), \;\;\;
\alpha = 
1,2. \eqno(3.10c) $$

We may verify that Eqs.(3.8) are fulfilled and also that the two
Casimir operators of the two subgroups are expressible by the
two Casimir operators of the group $ SO(1,3): $  $ M^2 =
\frac{1}{2} M^{ab} M_{ab} $ and $ \Gamma = \frac{-i}{3!}
\epsilon_{abcd} M^{ab}M^{cd} $

$$ (\tau^{\alpha }){ }^2 = \sum_{i=1}^{3} (\tau^{\alpha i}){ }^2 =
\frac{1}{4} \; [ \; M^2 + (-1)^{\alpha} \frac{3}{2} \Gamma \; ]
\;\;, \alpha = 1,2, \eqno(3.10d) $$

and oppositely.
These two invariant subalgebras are known as the $ SU(2)\times
SU(2) $ structure 
of the Lorentz group $ SO(1,3) $. While
 either  generators of 
spinorial  or of vectorial character fulfil
the same algebra, they
certainly have different representations, which we shall show in
Sect.4.

\vspace{3mm}

{ 3.1.2. \it Subgroups of SO(1,4). }

\vspace{3mm}

The interesting subalgebra of $ SO(1,4) $ is $ SO(1,3) $
discussed in the previous subsection, since it can be used to
describe spins of spinor, vector and scalar fields, that is
properties of  fields defined in the four dimensional part of the
space. If we choose $ M^{ab} $ with $ \; a,b \in \{ 0,1,2,3 \},
\; $ to form the subset of 
generators closing the algebra of $ \; SO(1,3) \;$, it is
meaningful to interpret the remaining generators of the group 
$ \; SO(1,4), \;\; $ that is $ M^{5a} , \; a \in \{ 0,1,2,3 \}
\; $, which do not form a subgroup, by a special name $
\gamma^a = -2i M^{5a} $. We showed
in ref. \cite{man2} that in the case of generators of spinorial
character $ \tilde{\gamma}^a = -2i \tilde S ^{5a}$ or $ \tilde
{\tilde \gamma} { }^a = -2i \tilde{\tilde S}{ }^{5a} $, these
generators may be recognized as the Dirac $\gamma^a  $ matrices,
with all the desired properties.

It is easy to show that  the two invariants of the subalgebra $
SO(1,3) $,  defined in the previous subsection  (we
shall call them $ M^{(4)}{ }^2 $ and $\Gamma^{(4)} $) can be
written as: 

$$ M^{(4)}{ }^2 = \frac{1}{2} M^{ab} M_{ab},\;\Gamma^{(4)} =  
i \frac{( -2i)^2}{4!} \epsilon_{abcd} M^{ab} M^{cd} 
= \frac{1}{4!} \epsilon_{abcd} \gamma^a \gamma^b \gamma^c
\gamma^d, $$

$$ a,b,c,d \in \{ 0,1,2,3 \}.  $$ 

This is valid for either of three cases; for 
$ M^{ab} = S^{ab} $ or $ M^{ab} = \tilde S^{ab} $ or $ M^{ab} =
\tilde{\tilde S}{ }^{ab} $. Since $ \{ M^{ab},
\gamma^d \} \neq 0 \;\; $,  $ SO(1,3) $ is not the
invariant subalgebra of $ SO(1,4). $  

While $ \frac{1}{2} M^{ab} M_{ab}, \;ab\in\{ 0,1,2,3,5 \} $ is the
invariant of the group $ SO(1,4) $, as well as of the subgroup
$SO(1,3) $ , the second invariant $\Gamma $ of the group $
SO(1,4) $ cannot be defined since  $d $ is odd.

\vspace{3mm}

{ 3.1.3. \it Subgroups of SO(6). }

\vspace{3mm}

The algebra of this group is isomorphic with the algebra of the
group $ SU(4) $, since both have the same number of generators
and the generators of $ SU(4) $ may, according to Eqs.(3.8) and
Eqs.(3.9), be expressed as  superpositions of $ M^{ab} $, with
the structure constants $ f^{1 ijk} $ presented in ref.
\cite{grein}.

This group contains as a subgroup the group $ SU(3). $
The structure constants of the  $ SU(3)$ algebra which can be found
in several text-books, are presented in Table I, 
while the coefficients $ {\it c }{ }^{1i}{ }_{ab}
$,  defining the generators of the subalgebra in terms of the
Lorentz ganerators $M^{ab}$,  are presented in Table II.

\begin{center}
\begin{tabular}{|c||c|c|c|c|c|c|c|c|c|}
\hline
$i $ & $1$ & $1$ & $1$ & $2$ & $2$ & $3$ & $3$ & $4$ & $6$ \\
\hline
$j $ & $2$ & $4$ & $5$ & $4$ & $5$ & $4$ & $6$ & $5$ & $7$ \\
\hline
$k $ & $3$ & $7$ & $6$ & $6$ & $7$ & $5$ & $7$ & $8$ & $8$ \\
\hline\hline
$f^{1ijk} $ & $1$ & $\frac{1}{2} $& $-\frac{1}{2}$ & $\frac{1}{2}$
& $\frac{1}{2}$ & $\frac{1}{2} $& $-\frac{1}{2}$ & $\frac{\sqrt{3}}{2}$
& $\frac{\sqrt{3}}{2}$\\
\hline
\end{tabular}

\end{center}
{ Table I: Structure constants \cite{grein} $f^{1ijk}$ of the
group $SU(3)$.  The coefficients $f^{1ijk}$ are antisymmetric
with respect to indices i,j,k.}

\vspace{3mm}

\begin{center}
\begin{tabular}{|c||c|c|c|c|c|c|c|c|c|c|c|c|c|c|c|c|c|}
\hline
$i $ & $1$ & $1$ & $2$ & $2$ & $3$ &
$3$ & $4$ & $4$ & $5$ & $5$ & $6$ & $6$ & $7$ & $7$ & $8$ & $8$ & $8$ \\
\hline
$a $ & $1$ & $2 $& $1$ &$2$&$1 $&$3$ & $2$ &
$1$ & $1$ & $2$ & $4$ & $3$ & $3$ &$4$ & $1$ & $3$ & $5$ \\
\hline
$b $ & $4$ & $3 $& $3$ &$4$&$2 $&$4$ & $5$ &
$6$ & $5$ & $6$ & $5$ & $6$ & $5$ &$6$ & $2$ & $4$ & $6$ \\
\hline\hline
$c^{1i}{ }_{ab} $ & $\frac{1}{2}$ & $-\frac{1}{2} $&
$\frac{1}{2}$ & $\frac{1}{2}$ 
& $\frac{1}{2}$ & $-\frac{1}{2} $& $\frac{-1}{2}$ &
$\frac{1}{2}$ 
& $\frac{1}{2}$ & $\frac{1}{2} $& $-\frac{1}{2}$ & $\frac{1}{2}$
& $\frac{1}{2}$ & $\frac{1}{2} $& $\frac{1}{2 {\sqrt 3}}$ &
$\frac{1}{2 \sqrt{3}}$ 
& $-\frac{1}{\sqrt{3}}$\\
\hline
\end{tabular}

\end{center}
{Table II: Table of coefficients $c^{1i}{ }_{ab} $ which
determine the generators 
of the subalgebra $SU(3)$ in terms of the generators $M^{ab}$ 
forming the Lorentz algebra of $SO(6)$ 
(Eq. $(3.8a))$. The coefficients have the property $c^{1i}{ }_{ab} = - 
c^{1i}{ }_{ba}$,  
$i \in$ $a,b \in \{1,8\}$. 
Only nonvanishing coefficients are presented.}

\vspace{3mm}

We may write, according to Eq.(3.8a) and Table II,

$$ \tau_{1} := \frac{1}{2}\; (M_{14} - M_{23}),\;\;\;\;
   \tau_{2} := \frac{1}{2}\; (M_{13} + M_{24}),\;\;\;\;
   \tau_{3} := \frac{1}{2}\; (M_{12} - M_{34}),$$
$$ \tau_{4} := \frac{1}{2}\; (M_{16} - M_{25}),\;\;\;\;
   \tau_{5} := \frac{1}{2}\; (M_{15} + M_{26}),\;\;\;\;
   \tau_{6} := \frac{1}{2}\; (M_{36} - M_{45}),$$
$$ \tau_{7} := \frac{1}{2}\; (M_{35} + M_{46}),\;\;\;\; 
   \tau_{8} := \frac{1}{2{\sqrt 3}}\; (M_{12} + M_{34} - 2 M_{56}).
\eqno(3.11) $$

The two invariants of the group $ SO(6): $ $\; M^2 = \frac{1}{2}
M^{ab} M_{ab}\; $ and $ \; \Gamma = $ $\frac{i(- 2i)^3}{6!} \;$ $
\epsilon_{abcdef}$ $  M^{ab} M^{cd} M^{ef} $ are at the same
time the two invariants 
of the subgroup $ SU(3) $. The two Casimir operators of the
group $SU(3)$ \cite{grein}   ${\it C}^1 = \frac {-2i}{3} \sum_{i,j,k}
f^{1ijk} \tau_i \tau_j \tau_k $ and $ {\it C}^2 = \sum_{ijk}
d^{1ijk} \tau_i \tau_j \tau_k $ 
can be expressed in terms of operators of the Lorentz
transformations forming the group $SO(6)$. We find for ${\it C}^1$

$$ {\it C^1} = \frac{1}{4} \{ \frac{1}{2} M^{ab} M_{ab} +
\frac{1}{3} ( M_{12} + M_{34} + M_{56} )^2 $$
$$- 2 ( M_{12} M_{34} - M_{13} M_{24} + M_{14} M_{23} )               
- 2 ( M_{12} M_{56} - M_{15} M_{26} + M_{16} M_{25} )$$
$$- 2 ( M_{34} M_{56} - M_{35} M_{46} + M_{36} M_{45} ) \}.
\eqno(3.12a)$$

The above relations are valid for generators of 
spinorial ($ \tilde S^{ab}, \tilde{ 
\tilde S}{ }^{ab} $) as well
as of vectorial ($ S^{ab} $) character.

In the case that operators of spinorial character $\tilde
S^{ab}$ are concerned, eq.(3.12a) simplifies to

$$ { \tilde{C}^1} = 1 + \frac{1}{3} \{ \tilde a_1 \tilde a_2
\tilde a_3 \tilde a_4 + \tilde a_1 \tilde a_2
\tilde a_5 \tilde a_6 + \tilde a_3 \tilde a_4
\tilde a_5 \tilde a_6 \}  $$

$$ \;\;\;\; = 1 + \frac{1}{3} \{ \tilde{\gamma}_1 \tilde{\gamma}_2
\tilde{\gamma}_3 \tilde{\gamma}_4 + \tilde{\gamma}_1
\tilde{\gamma}_2 
\tilde{\gamma}_5 \tilde{\gamma}_6 + \tilde{\gamma}_3
\tilde{\gamma}_4 
\tilde{\gamma}_5 \tilde{\gamma}_6 \}. $$

The general expression for the operator ${\it C}^2$ is much
longer. We shall present it here only for the case of spinorial
character, where it simplifies considerably. Using the
coefficients $ d^1{ }_{ijk} $ presented in Table III, we find


$$ { \tilde{C}^2} = i  \frac{5}{6} [ \tilde a_{1} \tilde a_{2}
\tilde a_{3} \tilde a_{4} \tilde a_{5} \tilde a_{6}  
+ \frac{1}{3} (\tilde a_{1} \tilde a_{2} + 
 a_{3} \tilde a_{4} +  \tilde a_{5} \tilde a_{6})]$$
$$ \;\;\;\; = i \frac{5}{6} [ \tilde{\gamma}_1  \tilde{\gamma}_2
\tilde{\gamma}_3 \tilde{\gamma}_4 \tilde{\gamma}_5
\tilde{\gamma}_6 + \frac{1}{3}( \tilde{\gamma}_1
\tilde{\gamma}_2 +  \tilde{\gamma}_3  \tilde{\gamma}_4 + 
 \tilde{\gamma}_5  \tilde{\gamma}_6)].\eqno(3.12b) $$

\begin{center}
\begin{tabular}{|c||c|c|c|c|c|c|c|c|c|c|c|c|c|c|c|c|}
\hline
$i $ & $1$ & $1$ & $1$ & $2$ & $2$ & $2$ & $3$ & $3$ & $3$ 
& $3$ & $3$ & $4$ & $5$ & $6$ & $7$  & $8$\\
\hline
$j $ & $1$ & $4$ & $5$ & $2$ & $4$ & $5$ & $3$ & $4$ & $5$ 
& $6$ & $7$ & $4$ & $5$ & $6$ & $7$  & $8$\\
\hline
$k $ & $8$ & $6$ & $7$ & $8$ & $7$ & $6$ & $8$ & $4$ & $5$ 
& $6$ & $7$ & $8$ & $8$ & $8$ & $8$  & $8$\\
\hline\hline
$d^{1ijk} $ & $\frac{1}{\sqrt{3}}$ & $\frac{1}{2} $& $\frac{1}{2}$ 
& $\frac{1}{\sqrt{3}}$ & $-\frac{1}{2} $& $\frac{1}{2}$ 
& $\frac{1}{\sqrt{3}}$ & $\frac{1}{2} $& $\frac{1}{2}$ 
& $-\frac{1}{2} $& $-\frac{1}{2}$ 
& $-\frac{1}{2 \sqrt{3}}$ & $-\frac{1}{2 \sqrt{3}}$
& $-\frac{1}{2 \sqrt{3}}$ & $-\frac{1}{2 \sqrt{3}}$ &
$-\frac{1}{ \sqrt{3}}$\\ 
\hline
\end{tabular}

\end{center}
{ Table III: The nonvanishing coefficients \cite{grein} $d^{1}{
}_{ijk}$, which determine the second Casimir operator
of the group $SU(3)$. The matrix of coefficients  $d^{1}{ }_{ijk}$  
is totally symmetric.}

\vspace{3mm}

One obtains equivalent relations for $ {\tilde{\tilde{C}}{ }^1}$
and $ {\tilde{\tilde{C}}{ }^2}$.

\vspace{3mm}

{ 3.1.4. \it Subgroups of SO(1,7). }

\vspace{3mm}

We present here as an instructive example the algebra of the group $
SO(1,7)$, which contains as a subgroup $ SO(1,4)\times SU(2)$. 
The generators $ M^{ab} $, with $ a,b \in \{0,1,2,3,5 \}, $ form the
subalgebra of the group $ SO(1,4) $ presented in Sect.
3.1.2., while $ M^{ab}$, with $ a,b \in \{ 6,7,8 \}, $ form the
algebra of $ SU(2) $, isomorphic to the algebra of $ SO(3) $,
presented in Sect.3.1.1. We shall use  these
subalgebras in Sect.4
to comment on the representations, the spinorial and the
vectorial, describing fields with spins and weak charges.
Again $ M^2 $ and $ \Gamma $ are the two Casimir operators of
the  group $ SO(1,7)$.

\vspace{3mm}

{ 3.1.5. \it Subgroups of SO(10). }

\vspace{3mm}

We choose this group since it contains as a subgroup the group
$ SU(5) $ which contains $ SU(3)\times
SU(2)\times U(1) $.

The coefficients ${ \it c}^{0i}{ }_{ab} , i \in \{1,24 \} , a,b
\in \{ 1,10 \}, $ which according to Eq. (3.8a) determine the
operators $ \tau^{0i} $ , defining the group $ SU(5) $, are
presented in Table VI in such a way that operators $ \tau^{0i} $  
demonstrate subalgebras
of $ SU(3) , SU(2) $ and $ U(1) $. We see that if we define

$$ \tau ^{1i}: = \tau ^{0 i+15} = {\it c}^{0 i+15}{ }_{ab}
M^{ab}, i \in \{ 1,8 \}, a,b \in \{ 1,10 \}, $$

$$ \tau ^{2i}: = \tau ^{0 i+12} = {\it c}^{0 i+12}{ }_{ab}
M^{ab}, i \in \{ 1,3 \}, a,b \in \{ 1,10 \}, \eqno (3.13) $$

$$ \tau ^{3i}: = \tau ^{0 i+23} = {\it c}^{0 i+23}{ }_{ab}
M^{ab}, i =1,  a,b \in \{ 1,10 \}, $$

operators $\tau^{1i}, \tau^{2i}, \tau^{3i} $ close the
subalgebras according to the following equations 

$$ \{ \tau^{1i}, \tau^{1j} \} = i f^{1ijk} \tau^{1k},\;\;\;
 \{ \tau^{2i}, \tau^{2j} \} = i \epsilon ^{ijk} \tau^{2k},\;\;\;
 \{ \tau^{\alpha i}, \tau^{\beta j} \} = 0,\;\; \alpha \neq \beta.
\eqno(3.13a) $$

Coefficients $ f^{1ijk} $ and $\epsilon^{ijk} $ are
structure constants of the groups $ SU(3) $ and $ SU(2) $,
respectively.



\begin{center}
\begin{tabular}{|c||c|c|c|c|c|c|c|c|c|c|c|c|c|c|c|c|}
\hline
$i $ & $1$ & $1$ & $2$ & $2$ & $3$ & $3$ & $4$ & $4$ & $5$ 
& $5$ & $6$ & $6$ & $7$ & $7$ & $8$  & $8$\\
\hline
$a $ & $1$ & $2$ & $1$ & $2$ & $1$ & $2$ & $1$ & $2$ & $3$ 
& $4$ & $3$ & $4$ & $3$ & $4$ & $3$  & $4$\\
\hline
$b $ & $8$ & $7$ & $7$ & $8$ & $10$ & $9$ & $9$ & $10$ & $8$ 
& $7$ & $7$ & $8$ & $10$ & $9$ & $9$  & $10$\\
\hline\hline
$c^{0i}{ }_{ab} $ & $\frac{1}{2}$ & $-\frac{1}{2} $& $\frac{1}{2}$ 
& $\frac{1}{2}$ & $\frac{1}{2} $& $-\frac{1}{2}$ 
& $\frac{1}{2}$ & $\frac{1}{2} $& $\frac{1}{2}$ 
& $-\frac{1}{2} $& $\frac{1}{2}$ 
& $\frac{1}{2}$ & $\frac{1}{2}$
& $-\frac{1}{2}$ & $\frac{1}{2}$ & $\frac{1}{2}$\\
\hline
\end{tabular}
\end{center}

\begin{center}
\begin{tabular}{|c||c|c|c|c|c|c|c|c|c|c|c|c|c|c|}
\hline
$i $ & $9$ & $9$ & $10$ & $10$ & $11$ & $11$ & $12$ & $12$ & $13$ 
& $13$ & $14$ & $14$ & $15$ & $15$ \\
\hline
$a $ & $5$ & $6$ & $5$ & $6$ & $5$ & $6$ & $5$ & $6$ & $7$ 
& $8$ & $7$ & $8$ & $7$ & $9$\\
\hline
$b $ & $8$ & $7$ & $7$ & $8$ & $10$ & $9$ & $9$ & $10$ & $10$ 
& $9$ & $9$ & $10$ & $8$ & $10$\\
\hline\hline
$c^{0i}{ }_{ab} $ & $\frac{1}{2}$ & $-\frac{1}{2} $& $\frac{1}{2}$ 
& $\frac{1}{2}$ & $\frac{1}{2} $& $-\frac{1}{2}$ 
& $\frac{1}{2}$ & $\frac{1}{2} $& $\frac{1}{2}$ 
& $-\frac{1}{2} $& $\frac{1}{2}$ 
& $\frac{1}{2}$ & $\frac{1}{2}$
& $-\frac{1}{2}$\\
\hline
\end{tabular}
\end{center}

\begin{center}
\begin{tabular}{|c||c|c|c|c|c|c|c|c|c|c|c|c|c|c|c|c|c|}
\hline
$i $ & $16$ & $16$ & $17$ & $17$ & $18$ & $18$ & $19$ & $19$ & $20$ 
& $20$ & $21$ & $21$ & $22$ & $22$ & $23$  & $23$ & $23$\\
\hline
$a $ & $1$ & $2$ & $1$ & $2$ & $1$ & $3$ & $1$ & $2$ & $1$ 
& $2$ & $4$ & $3$ & $3$ & $4$ & $1$  & $3$ &$5$\\
\hline
$b $ & $4$ & $3$ & $3$ & $4$ & $2$ & $4$ & $6$ & $5$ & $5$ 
& $6$ & $5$ & $6$ & $5$ & $6$ & $2$  & $4$ & $6$\\
\hline\hline
$c^{0i}{ }_{ab} $ & $\frac{1}{2}$ & $-\frac{1}{2} $&
$\frac{1}{2}$ 
& $\frac{1}{2}$ & $\frac{1}{2} $& $-\frac{1}{2}$ 
& $\frac{1}{2}$ & $-\frac{1}{2} $& $\frac{1}{2}$ 
& $\frac{1}{2} $& $-\frac{1}{2}$ 
& $\frac{1}{2}$ & $\frac{1}{2}$
& $\frac{1}{2}$ & $\frac{1}{2 \sqrt{3}}$ & $\frac{1}{2
\sqrt{3}}$ 
& $-\frac{1}{\sqrt{3}}$\\
\hline
\end{tabular}
\end{center}
\begin{center}
\begin{tabular}{|c||c|c|c|c|c|}
\hline
$i $ & $24$ & $ 24$ & $ 24$  & $ 24$ & $24$  \\
\hline
$a $ & $ 1 $ & $ 3$ & $ 5 $ & $7$ & $ 9$ \\
\hline
$b $ & $2$ & $4$ & $ 6$ & $ 8$ & $ 10$ \\
\hline\hline
$c^{0i}{ }_{ab} $ & $-\frac{1}{\sqrt{15}}$ &
$-\frac{1}{\sqrt{15}}$ & 
$-\frac{1}{\sqrt{15}}$ & $\sqrt{\frac{3}{20}}$ &
$\sqrt{\frac{3}{20}}$ \\ 
\hline
\end{tabular}
\end{center}
{ Table IV: Table of  coefficients $c^{0i}{ }_{ab}$,
$c^{0i}{ }_{ab} =- c^{0i}{ }_{ba}$, $i \in \{1,24\}$ 
$a,b \in \{1,10\}$ determining the operators $\tau^{0i} =
c^{0i}{ }_{ab} M^{ab}$, 
which form the algebra of $SU(5)$. 
According to Eq. (3.13) operators $\tau^{0i}$ 
demonstrate the structure 
 $SU(3)\times SU(2) \times U(1)$.}

\vspace{3mm}

From  Table IV and Eq.(3.8a) we find the  expressions for
$ \tau ^{1 i} $ forming the algebra of $SU(3)$, they  are
presented in Eq.(3.11),
the expressions

$$ \tau^{21}: =  \frac{1}{2} \;(M_{7 10} - M_{8 9}),\;\;\;\;
   \tau^{22}: =  \frac{1}{2} \;(M_{7 9} + M_{8 10}),\;\;\;\;
   \tau^{23}: =  \frac{1}{2} \;(M_{7 8} - M_{9 10}), \eqno(3.14a)$$

forming the algebra of $SU(2)$ and the expression

$$ \tau^{31}: = \sqrt{\frac{3}{5}}\;\; [- \frac{1}{3}\; (M_{1 2} +
M_{3 4} + M_{5 6}) + \frac{1}{2} \;(M_{7 8} + M_{9 10})], 
\eqno(3.14b)$$ 

forming the algabra $U(1)$.

Operators $ \tau^{\alpha i} $ define either spinorial ( if $
M^{ab} = \tilde S^{ab} ( \tilde {\tilde S}{ }^{ab} ) $) or vectorial
( if $ M^{ab} = S^{ab} $) representations.

\vspace{3mm}

{3.1.6. \it Subgroups of SO(1,14). }

\vspace{3mm}

The subalgebra of $ SO(1,4)\times SO(10) $ of the algebra
$SO(1,14)$ may be 
used to describe spins $ ( SO(1,4) ) $ and charges $( SO(10)) $ of
fields. Both subalgebras were discussed in previous subsections,
3.1.2 and 3.1.5, respectively.

The choice of the group $ SO(1,4) $ rather than $SO(1,3)$
enables us to define generators $ \gamma ^a = -2i M^{5a} $ as we
know from Sect.3.1.2. We can then write for one of the
invariants of the subgroups $ SO(1,3) $ and $ SO(10) $, 
respectively : $ \Gamma^{(4)} = i \frac{1}{4!} \epsilon_{abcd}
\gamma^a \gamma^b \gamma^c \gamma^d,\;\; a,b,c,d \in {0,1,2,3,}
$ and $ \Gamma^{(10)} = i \frac{1}{10!}$ $ \epsilon_{a_1 a_2
....a_{10}} \gamma^{a_1}$ $ \gamma^{a_2} ....\gamma^{a_{10}} $.
This is true for  operators of either spinorial or vectorial
character.

We shall comment on spinorial and vectorial representations of
these subalgebras in Sect.4.

\vspace{4mm}

{ 4. \bf Spinorial and vectorial representations of
 $SO(d)$ or $SO(1,d-1)$ and of subgroups in 
Grassmann space}

\vspace{3mm}

 In this section we shall present some representations of
generators of the Lorentz transformations of spinorial
character $ \tilde S^{ab} $ and representations of operators of 
vectorial character $ S^{ab} $ for the groups discussed in the
previous section. The former
define representations which include what is called fundamental
representations, the 
latter define representations which include what is called
regular or adjoint representations.  We shall see that
one can find singlets of both types, of spinorial as well as of 
vectorial character as well. Representations of
operators $ \tilde{\tilde S}{ }^{ab} $ can be found in an
equivalent way to those  of $ \tilde S^{ab} $. 

The choice of  Grassmann space as the
coordinate space over which the space of vectors - the basis
for the generators of the Lorentz transformations - is spanned,
enables the unification of spin and charges, and limits the
vectors to those which correspond to spin $ 
\frac{1}{2} $, if operators of spinorial character are concerned
and to spin one or zero , if operators of vectorial character
are concerned. Spins
$ \frac{3}{2} $ and $2$ follow for  tensor fields \cite{man1,man2}.

According to Eq.(2.5) the vector space, spanned over the
d-dimensional Grassmann space, is finite dimensional. It has $
2^{d-1}$ vectors of an odd Grassmann 
character and $ 2^{d-1} $ vectors of an even Grassmann
character. 

 In this section  we shall present operators $ M^{ab} $ in the
coordinate representation, define the integral over the Grassmann
space \cite{berez}, the inner product of the vectors, as well
as irreducible 
representations of generators $ M^{ab}$ of spinorial and
vectorial character.

\vspace{3mm}

{\it 4.1. Integrals on Grassmann space. Inner products.}

\vspace{3mm}

We assume that differentials of Grassmann coordinates $ d\theta^a
$ fulfill the Grassmann anticommuting relations \cite{man2,berez}

$$ \{ d\theta^a, d\theta^b \} = 0  \eqno(4.1)$$

and we introduce a single integral over the whole interval of
$d\theta^a $ 

$$ \int d\theta^a = 0, \;\; \int d\theta^a \theta^a = 1, a =
0,1,2,3,5,..,d, \eqno(4.2)$$

and the multiple integral over d coordinates

$$ \int d^d \theta  \theta^0 \theta^1 \theta^2 \theta^3
\theta^4...\theta^d = 1, \eqno(4.3) $$

with 

$$ d^d \theta: = d\theta^d...d\theta^3 d\theta^2 d\theta^1
d\theta^0. $$

We define \cite{man2} the inner product of two vectors $
<\varphi|\theta> $ and $ <\theta|\chi> $, with $
<\varphi|\theta> = <\theta|\varphi>^* $ as follows:

$$ <\varphi|\chi> = \int d^d\theta ( \omega<\varphi|\theta>)
<\theta| \chi>, \eqno(4.4) $$

with the weight function $\omega$ 

$$\omega = \prod_{k=0,1,2,3,..,d}
(\frac{\partial}{\partial \theta^k}  + \theta^k ),
\eqno(4.4a)$$ 

which operates on the first function $ <\varphi|\theta> $ only, 
and we define

$$ (\alpha^{a_1 a_2...a_k} \theta^{a_1}
\theta^{a_2}...\theta^{a_k})^{+} =
(\theta^{a_k}).....(\theta^{a_2}) 
(\theta^{a_1}) (\alpha^{a_1 a_2...a_k})^{*}.\eqno(4.4b)$$

Then $\theta^{a}{ }^* = \theta^{a},\;\;\theta^{a}{ }^+ = -
\eta^{aa} \frac{\partial}{\partial \theta^{a}} $ and
$\frac{\partial}{\partial \theta^{a}}{ }^+ = \eta^{aa}
\theta^{a}, $ while   $\tilde a^a{ }^+ =  \eta^{aa} \tilde a^a $ and
$\tilde{\tilde a}{ }^a{ }^+ =   \eta^{aa} \tilde{\tilde a}{ }^a $.  
Accordingly the generators of the Lorentz
transformations of Eqs.(3.1) are self adjoint $ ( if\; a \neq 0
\;\;and\;\; b \neq 0 )$ or anti self adjoint $(if\; a = 0\;\;
or\;\; b = 0 )$ operators. 

Either the volume element $ d^d\theta $ or the weight function $
\omega$ are invariants with respect to the Lorentz
transformations ( both are scalar densities of weight - 1).

\vspace{3mm}

{4.2. \it Explicit expressions for operators of the Lorentz
transformations in Grassmann space }

\vspace{3mm}

We express the generators of the Lorentz transformations in
terms of coordinates and left derivatives in order to be able to
solve the eigenvalue problem in the coordinate representation.

According to Eqs.(2.7) and (3.1) we find

$$ S^{ab} = -i ( \theta^a \frac{\partial}{\partial \theta_b}
- \theta^b \frac{\partial}{\partial \theta_a} ), \eqno(4.5a) $$

$$ \tilde S ^{ab} = \frac{-i}{2}( \frac{\partial}{\partial
\theta_a} + 
\theta^a ) ( \frac{\partial}{\partial \theta_b} +
\theta^b ) ,\; if a \neq b, \; \eqno(4.5b) $$

$$ \tilde {\tilde S}{ } ^{ab} = \frac{i}{2}(
\frac{\partial}{\partial \theta_a} - 
\theta^a ) ( \frac{\partial}{\partial \theta_b} - 
\theta^b ) ,\; if a \neq b, \; \eqno(4.5c) $$

$$ \tilde a^a = (\frac{\partial}{\partial \theta_a} +
\theta^a ),\;\; \tilde{\tilde a}{ }^a = i
(\frac{\partial}{\partial \theta_a} - 
\theta^a ). \eqno(4.5d) $$

Accordingly, the superpositions of the operators of the Lorentz
transformations, as well as their invariants, can be expressed in
terms of $ M^{ab} $ from Eqs.(4.5).

\vspace{3mm}

{4.3.\it The eigenvalue problem}

\vspace{3mm}

To find the irreducible representations of the desired groups or
subgroups, we solve the eigenvalue problem for the commuting
operators which define the algebra of the group or subgroups 

$$ <\theta|\tilde A_i|\tilde{\varphi}> = \tilde{\alpha}_i
<\theta|\tilde{\varphi}> ,\;\;
<\theta|A_i|\varphi> = \alpha_i <\theta|\varphi>,\;\;i = \{1,r\}
,\eqno(4.6)$$ 

where $ \tilde{A}_i $ and $ A_i $ stand for $r$ commuting
operators 
of spinorial and vectorial character, respectively.

To solve  equations (4.6) we express the operators in the
coordinate representation (Eqs.(4.5)) and write the eigenvectors
as polynomials of $\theta^a$. We  orthonormalize the vectors
according to the inner product, defined in Eq.(4.4)

$$ <{ }^a \tilde{\varphi}_i |{ }^b \tilde{\varphi}_j> =
\delta^{ab} \delta_{ij},\;\;\; <{ }^a {\varphi}_i |{ }^b
{\varphi}_j> = \delta^{ab} \delta_{ij}, \eqno(4.6a)$$

where index $a$ distinguishes between vectors of different
irreducible representations and index j between vectors of the
same irreducible representation. Eq.(4.6a) determines the
orthonormalization condition for spinorial and vectorial
representations, respectively.

\vspace{3mm}

{4.3.1. \it Representations of the group $ SO(1,3) $ and 
subgroups}  

\vspace{3mm}

{\bf $SO(3)$}

\vspace{2mm}

We shall find the representations of the subgroup $ SO(3) $
first. According to Eqs. (3.6) and (3.10), we look for the
eigenvectors of the operators $ \tau^{1i}$, choosing $i = 3 $,
and $ (\tau^{1})^2 $. We find  $\tau^{1}{ }^3 = M^{12}$ and
 $ (\tau^{1})^2 = (M^{12})^2 +(-M^{13})^2 + (M^{23})^2  $.

Looking for solutions in the spinorial case we take operators of
Eq.(4.5b) and find eight vectors, which can be arranged into
{\it two bispinors} ( each bispinor has two vectors ) of an {\it
odd} and {\it two bispinors} of an 
{\it even} Grassmann
character. We present these eight vectors in Table Va, together
with the eigenvalues of the two operators $\tilde \tau^{13} $ and
$(\tilde\tau^{1})^2$.

\begin{center}
\begin{tabular}{|c|c||c|r|c|}
\hline
$a$ & $i$ & $<\theta|\tilde{\varphi}^{a}_{i}>$& 
$\tilde{\tau}^{31}$ & $(\tilde{\tau}^{1})^{2}$\\
\hline\hline 
$1$ & $1$ & $ \frac{1}{\sqrt{2}} (\theta^{1} - i \theta^{2})$& 
$\frac{1}{2}$ & $ \frac{3}{4}$ \\
\hline
$1$ & $2$ & $ -\frac{1}{\sqrt{2}} (1 +
i\theta^{1}\theta^{2})\theta^{3}$& 
$-\frac{1}{2}$ & $ \frac{3}{4}$ \\
\hline\hline
$2$ & $1$ & $ \frac{1}{\sqrt{2}} (1 -
i\theta^{1}\theta^{2})\theta^{3}$& 
$\frac{1}{2}$ & $ \frac{3}{4}$ \\
\hline
$2$ & $2$ & $ -\frac{1}{\sqrt{2}} (\theta^{1} + i \theta^{2})$& 
$-\frac{1}{2}$ & $ \frac{3}{4}$ \\
\hline\hline
$3$ & $1$ & $ \frac{1}{\sqrt{2}} (\theta^{1} -
i\theta^{2})\theta^{3}$& 
$\frac{1}{2}$ & $ \frac{3}{4}$ \\
\hline
$3$ & $2$ & $ -\frac{1}{\sqrt{2}} (1 + i\theta^{1}\theta^{2})$& 
$-\frac{1}{2}$ & $ \frac{3}{4}$ \\
\hline\hline
$4$ & $1$ & $ \frac{1}{\sqrt{2}} (1 - i\theta^{1}\theta^{2})$& 
$\frac{1}{2}$ & $ \frac{3}{4}$ \\
\hline
$4$ & $2$ & $ -\frac{1}{\sqrt{2}} (\theta^{1} + i
\theta^{2})\theta^{3}$& 
$-\frac{1}{2}$ & $ \frac{3}{4}$ \\
\hline
\end{tabular}
\end{center}
Table Va: Eigenvectors of  commuting operators of a 
spinorial character for the group $SO(3)$, arranged into four
irreducible 
representations, two of an odd and two of an even Grassmann
character.  
Eigenvalues of the operators $\tilde{\tau}^{13}$ and
$(\tilde{\tau}^{1})^{2}$ 
are also presented.
Vectors are orthonormalized according to the inner product 
defined in Eq. (4.4).

\vspace{3mm}

The operators $\tilde\tau^{1}{ }_{\pm} = \tilde\tau^{11} \pm i
\tilde\tau^{12}$ rotate one vector of a bispinor into another
vector 
of the same bispinor. We have, therefore, instead of one, four
independent bispinors. To describe fermions in a standard way,
the two 
bispinors of an
odd Grassmann character are needed. Bispinors in
Table Va are arranged in such a way that the matrix
representations of the operators $\tilde\tau^{1i},
\;i = 1,2,3 $ are for any bispinor just the 
Pauli $ \sigma_i $ matrices.

To look for the solutions in the vectorial case, we have to take
the operators of Eq.(4.5a). We find {\it two scalars} and {\it two
three vectors}. Again, one scalar and one three vector have
an{\it odd}, another scalar and another three vector have 
an{\it even} Grassmann character. These eight vectors are presented 
in Table Vb.

\begin{center}
\begin{tabular}{|c|c||c|r|c|}
\hline
$a$ & $i$ & $<\theta|{\varphi}^{a}_{i}>$& 
${\tau}_{13}$ & $({\tau}^{1})^{2}$\\
\hline\hline 
$1$ & $1$ & $ 1$& 
$ 0 $ & $ 0$ \\
\hline\hline
$2$ & $1$ & $ \frac{1}{\sqrt{2}} (\theta^{1} - i
\theta^{2})\theta^{3}$& 
$ 1$ & $ 2$ \\
\hline
$2$ & $2$ & $ - \theta^{1}\theta^{2}$& 
$0 $ & $ 2$ \\
\hline
$2$ & $3$ & $ -\frac{1}{\sqrt{2}} (\theta^{1} + i
\theta^{2})\theta^{3}$& 
$-1$ & $ 2$ \\
\hline\hline
$3$ & $3$ & $ \theta^{1} \theta^{2} \theta^{3}$ & $ 0$ & $ 0$ \\
\hline\hline
$4$ & $1$ & $ \frac{1}{\sqrt{2}} (\theta^{1} - i\theta^{2})$& 
$1$ & $ 2$ \\
\hline
$4$ & $2$ & $ -\theta^{3}$& 
$0$ & $ 2$ \\
\hline
$4$ & $3$ & $ -\frac{1}{\sqrt{2}} (\theta^{1} + i \theta^{2})$& 
$-1$ & $ 2$ \\
\hline
\end{tabular}
\end{center}

Table Vb: Eigenvectors of  commuting operators of a vectorial
character for the group $SO(3)$, arranged into two scalars and 
two three vectors. Half of the vectors have an even and half an
odd Grassmann character. 

\vspace{3mm}

Each of the three vectors are arranged in such a way that 
the matrix representations of the operators $\tau^{1i}$ 
are the usual $ 3 \times 3 $ matrices \cite{man2}. 
The three vectors are in the adjoint representations with
respect to the bispinors of Table Va. To describe
vectors and scalars only Grassmann even solutions are taken.

\vspace{2mm}

{\bf $SO(1,3)$ }

\vspace{2mm}

To find representations manifesting the $SU(2) \times SU(2)$
structure of the group $ 
SO(1,3)$, one has to solve the eigenvalue problem (4.6) for
each of the two types of  operators $\tau^{1i}$ and
$\tau^{2i},\; i = 1,2,3,\;$ defined by coefficients (3.10). We
gave the solutions in  ref. \cite{man2} for both types of 
operators. 

In the spinorial case, there are {\it eight bispinors},
forming the irreducible representations of the group $SO(1,3)$.
Half of them have an {\it odd} and half an {\it even} Grassmann
character. They can be further  distinguished with respect to the
eigenvalue of the operator $\tilde\Gamma $, which is either $+1$
or $-1$. We speak about chiral 
representations. They cannot be arranged into four four spinors
unless we define the Grassmann even operators $\tilde\gamma^a$ which 
connect the two Grassmann odd two bispinors into a four spinor.
This can be done \cite{man2} within the group $SO(1,4)$, which will
be discussed in the next subsection. 

In the  vectorial case we find {\it two scalars} and {\it two
three vectors} of an {\it even} Grassmann character and {\it two
four vectors} of an {\it odd} Grassmann character \cite{man2}.
Three vectors form the adjoint representations with respect to
bispinors due to the $ SU(2) \times SU(2) $ structure of the
group $ SO(1,3) $.

\vspace{3mm}

{4.3.2. \it Representations of the group $SO(1,4)$ and subgroups}

\vspace{3mm}

{\bf $SO(1,3)$ }

\vspace{3mm}

We see that the two sets of  Casimir operators, $ M^2$ and
$\Gamma$ of  Subsect.3.1.1. and $ M^{(4)}{ }^2 $ and $
\Gamma^{(4)} $ of Subsect.3.1.2., are equal since 
they are both defined in the vector space spanned over the
coordinates  $\theta^a$, $ a \in \{0,1,2,3\} $. 

We defined  in Subsect. 3.1.2. the operators which are not
included in the subgroup $SO(1,3)$,
as the $\gamma^a$ operators $\tilde{\gamma}^a =
-2i \tilde S^{5a} = - \tilde a^5 \tilde a^a $ for the spinorial
type of operators, and $\gamma^a = -2iS^{5a} $ for the vectorial
type of operators.

We look for the representations \cite{man2} of the subalgebras $
SU(2) \times SU(2) $ (Eq.3.10b) in the space spanned over the
five dimensional coordinate space: $16$ vectors have an odd and
$16$ vectors an even Grassmann character. 
Solving the eigenvalue problem for the operators of spinorial
character  within the vector space of an
odd Grassmann character, we find {\it eight bispinors}, which we
arrange into four four spinors such 
that a bispinor which is a singlet with respect to one of the
two $SU(2)$ subgroups and a doublet with respect to the other
$SU(2)$ and a bispinor which is the doublet with repect to the
first $SU(2)$ group and the singlet with respect to the second
$SU(2)$, are transformed into each other by operators $\tilde
\gamma^a$. Each bispinor forms the irreducible
representation with respect to the generators of the Lorentz
transformations in the four dimensional subspace of the five
dimensional Grassmann space.  We present these vectors, 
taken from ref.\cite{man2},  in Table VIa. One can find the matrix
representations of the operators in the same reference.

\begin{center}
\begin{tabular}{|c|c||c||r|r|r|r|c|r|c|}
\hline
$ a$& $i$ & 
$<\theta|\tilde{\varphi}^{a}_{i}>$ &  $\tilde{S}_{3}$ & 
$\tilde{K}_{3}$ & $ \tilde{\Gamma} $&
$\tilde{\tau}^{13}$ &  $(\tilde{\tau}^{1})^{2}$ & 
$\tilde{\tau}^{23}$ & $(\tilde{\tau}^{2})^{2}$ \\
\hline
\hline 
$1$ & $1$ & $ \frac{1}{2}
( \theta^{1} - i \theta^{2})(\theta^{0} - \theta^{3})\theta^{5}$
& 
$ \frac{1}{2}$ & $ \frac{i}{2}$ &$ 1$ & $0$ & $ 0 $ 
&$ \frac{1}{2} $ & $\frac{3}{4}$ \\
\hline
$1$ & $2$ & $ -\frac{1}{2}(1 + 
i \theta^{1}\theta^{2})(1 - \theta^{0}\theta^{3})\theta^{5}$ &
$ -\frac{1}{2}$ & $ -\frac{i}{2}$ &$ 1$ & $0$ & $ 0 $ 
&$ -\frac{1}{2} $ & $\frac{3}{4}$ \\
\hline
$2$ & $1$ & $ -\frac{1}{2}
(\theta^{1} - i \theta^{2})(1 - \theta^{0}\theta^{3})$ & 
$ \frac{1}{2}$ & $- \frac{i}{2}$ &$ -1$  
&$ \frac{1}{2} $ & $\frac{3}{4}$ & $0$ & $ 0$ \\
\hline
$2$ & $2$ & $ -\frac{1}{2}(1 + 
i \theta^{1}\theta^{2})(\theta^{0} - \theta^{3})$ &
$ -\frac{1}{2}$ & $ \frac{i}{2}$ &$ -1$ 
&$ -\frac{1}{2} $ & $\frac{3}{4}$ & $0$ & $0 $ \\
\hline\hline 
$3$ & $1$ & $ \frac{1}{2}
( \theta^{1} - i \theta^{2})(1 + \theta^{0}\theta^{3})$ & 
$ \frac{1}{2}$ & $ \frac{i}{2}$ &$ 1$ & $0$ & $ 0 $ 
&$ \frac{1}{2} $ & $\frac{3}{4}$ \\
\hline
$3$ & $2$ & $ -\frac{1}{2}(1 + 
i \theta^{1}\theta^{2})(\theta^{0} + \theta^{3})$ &
$ -\frac{1}{2}$ & $ -\frac{i}{2}$ &$ 1$ & $ 0$ & $0$ &
$ -\frac{1}{2} $ & $\frac{3}{4}$ \\
\hline
$4$ & $1$ & $ \frac{1}{2}
( \theta^{1} - i \theta^{2})(\theta^{0} + \theta^{3})\theta^{5}$
& 
$ \frac{1}{2}$ & $- \frac{i}{2}$ &$ -1$ 
&$ \frac{1}{2} $ & $\frac{3}{4}$& $ 0 $ & $ 0$  \\
\hline
$4$ & $2$ & $ -\frac{1}{2}(1 + 
i \theta^{1}\theta^{2})(1 + \theta^{0}\theta^{3})\theta^{5}$ &
$ -\frac{1}{2}$ & $ \frac{i}{2}$ &$ -1$ 
&$ -\frac{1}{2} $ & $\frac{3}{4}$ & $ 0 $ & $ 0 $\\
\hline\hline
$5$ & $1$ & $ \frac{1}{2}(1 -
i \theta^{1}\theta^{2})(\theta^{0} - \theta^{3})$ &
$ \frac{1}{2}$ & $ \frac{i}{2}$ &$ 1$ & $ 0$ & $ 0$  
&$ \frac{1}{2} $ & $\frac{3}{4}$ \\
\hline
$5$ & $2$ & $ -\frac{1}{2}(\theta^{1} + i \theta^{2})(1 -
\theta^{0}\theta^{3})$ & 
$ -\frac{1}{2}$ & $ -\frac{i}{2}$ &$ 1$ & $ 0$ & $0 $
&$ -\frac{1}{2} $ & $\frac{3}{4}$\\
\hline
$6$ & $1$ & $ \frac{1}{2}(1 - i \theta^{1}\theta^{2})(1 -
\theta^{0}\theta^{3})\theta^{5}$ & 
$ \frac{1}{2}$ & $ -\frac{i}{2}$ &$ -1$ 
&$ \frac{1}{2} $ & $\frac{3}{4}$ & $ 0 $ & $ 0 $\\
\hline
$6$ & $2$ & $ \frac{1}{2}(\theta^{1} + i \theta^{2})(\theta^{0}
- \theta^{3})\theta^{5}$ & 
$ -\frac{1}{2}$ & $ \frac{i}{2}$ &$ -1$ 
&$ -\frac{1}{2} $ & $\frac{3}{4}$ & $ 0 $ & $ 0 $\\
\hline
\hline
$7$ & $1$ & $ \frac{1}{2}(1 -
i \theta^{1}\theta^{2})(1 + \theta^{0}\theta^{3})\theta^{5}$ &
$ \frac{1}{2}$ & $ \frac{i}{2}$ &$ 1$ & $ 0$ & $ 0$  
&$ \frac{1}{2} $ & $\frac{3}{4}$ \\
\hline
$7$ & $2$ & $ -\frac{1}{2}(\theta^{1} + i \theta^{2})(\theta^{0}
+ \theta^{3})\theta^{5}$ & 
$ -\frac{1}{2}$ & $ -\frac{i}{2}$ &$ 1$ & $ 0$ & $0 $
&$ -\frac{1}{2} $ & $\frac{3}{4}$\\
\hline
$8$ & $1$ & $ -\frac{1}{2}(1 - i \theta^{1}
\theta^{2})(\theta^{0} + \theta^{3})$ & 
$ \frac{1}{2}$ & $ -\frac{i}{2}$ &$ -1$
& $ \frac{1}{2} $ & $\frac{3}{4}$ & $ 0$ & $ 0$ \\
\hline
$8$ & $2$ & $ \frac{1}{2}(\theta^{1} + i \theta^{2})(1 +
\theta^{0}\theta^{3})$ & 
$ -\frac{1}{2}$ & $ \frac{i}{2}$ &$ -1$ 
&$ -\frac{1}{2} $ & $\frac{3}{4}$ & $ 0 $ & $ 0 $\\
\hline
\end{tabular}
\end{center}

Table VIa: Irreducible representations of the two subgroups
$SU(2) \times SU(2) $ embedded into the group $SO(1,4)$ as
defined by the generators of spinorial character. There are
eight bispinors, two by two bispinors connected into a four
spinor by the operators $ \tilde \gamma^a $ in the way that
$\tilde \gamma^0$ is diagonal. We present
eigenvalues of the commuting operators $ \tilde S_3 = \tilde
S^{12}, \tilde K_3 = \tilde S^{03}, \tilde
{\Gamma}^{(4)}, \tilde {\tau}^{13}, (\tilde \tau^1)^2, \tilde
\tau^{23}, (\tilde \tau^2)^2 $. 

\vspace{3mm}

Analyzing the space of $16$ vectors of an even Grassmann
character with respect to the operators of vectorial character,
we find {\it two scalars} and {\it two three vectors} which do
not depend on the coordinate $\theta^5$ and {\it two four
vectors} which do depend on the coordinate $\theta^5$. They are
vectors and scalars with respect to the operators of the Lorentz
transformations in the four dimensional subspace of the five
dimensional space. 
We present these vectors, taken from ref. \cite{man2}, in Table
VIb. One can find in the above reference also the matrix
representation of the 
operators. 

\vspace{3cm}

\begin{center}
\begin{tabular}{|c|c||c||c|r|r|r|r|c|r|c|}
\hline
$ a$& $i$ & 
$<\theta|{\varphi}^{a}_{i}>$ & ${S}^{2}$ &  ${S}_{3}$ & 
${K}_{3}$ & $3 \frac{\Gamma}{8} $ &
${\tau}^{13}$ &  $({\tau}^{1})^{2}$ & 
${\tau}^{23}$ & $({\tau}^{2})^{2}$ \\
\hline
\hline 
$1$ & $1$ & $ (\frac{1}{{\sqrt 2}})(1 + 
i \theta^{0}\theta^{1}\theta^{2}\theta^{3})$ & $ 0$ & $ 0$ & $
0$ & $0$ & $ 0 $ 
&$ 0 $ & $0$ & $0$ \\
\hline
$2$ & $1$ & $ (\frac{1}{{\sqrt 2}})(1 - 
i \theta^{0}\theta^{1}\theta^{2}\theta^{3})$ & $ 0$ & $ 0$ & $
0$ & $0$ & $ 0 $ 
&$ 0 $ & $0$ & $0$ \\
\hline\hline
$3$ & $1$ & $ \frac{1}{2}(\theta^{0} - \theta^{3})(\theta^{1} -
i \theta^{2})$ & 
$ 1$ & $ 1$ & $ i$ & $1$ & $ 0 $ 
&$ 0 $ & $1$ & $2$ \\
\hline
$3$ & $2$ & $- \frac{1}{{\sqrt 2}}(i\theta^{1}\theta^{2} -
\theta^{0} \theta^{3})$ & 
$ 1$ & $ 0$ & $ 0$ & $1$ & $ 0 $ 
&$ 0 $ & $0$ & $2$ \\
\hline
$3$ & $3$ & $ -\frac{1}{2}(\theta^{0} + \theta^{3})(\theta^{1} +
i \theta^{2})$ & 
$ 1$ & $ -1$ & $ -i$ & $1$ & $ 0 $ 
&$ 0 $ & $-1$ & $2$ \\
\hline\hline

$4$ & $1$ & $ \frac{1}{2}(\theta^{0} + \theta^{3})(\theta^{1} -
i \theta^{2})$ & 
$ 1$ & $ 1$ & $ -i$ & $-1$ & $ 1 $ 
&$ 2 $ & $0$ & $0$ \\
\hline
$4$ & $2$ & $- \frac{1}{{\sqrt 2}}(i\theta^{1}\theta^{2} +
\theta^{0} \theta^{3})$ & 
$ 1$ & $ 0$ & $ 0$ & $-1$ & $ 0 $ 
&$ 2 $ & $0$ & $0$ \\
\hline
$4$ & $3$ & $ -\frac{1}{2}(\theta^{0} - \theta^{3})(\theta^{1} +
i \theta^{2})$ & 
$ 1$ & $ -1$ & $ i$ & $-1$ & $ -1 $ 
&$ 2 $ & $0$ & $0$ \\
\hline\hline
$5$ & $1$ & $ \frac{1}{2}(1 + \theta^{0}\theta^{3})(\theta^{1} -
i \theta^{2}) \theta^5$ & 
$ \frac{3}{4}$ & $ 1$ & $ 0$ & $0$ & $\frac{1}{2} $ 
&$ \frac{3}{4} $ & $\frac{1}{2}$ & $\frac{3}{4}$ \\
\hline
$5$ & $2$ & $ \frac{1}{2}(\theta^{0} - \theta^{3})(1 -
i\theta^{1} \theta^{2}) \theta^5$ & 
$ \frac{3}{4}$ & $ 0$ & $ i$ & $0$ & $-\frac{1}{2} $ 
&$ \frac{3}{4} $ & $\frac{1}{2}$ & $\frac{3}{4}$ \\
\hline
$5$ & $3$ & $ \frac{1}{2}(1 - \theta^{0} \theta^{3})(\theta^{1}
+ i \theta^{2}) \theta^5$ & 
$ \frac{3}{4}$ & $ -1$ & $ 0$ & $0$ & $-\frac{1}{2} $ 
&$ \frac{3}{4} $ & $-\frac{1}{2}$ & $\frac{3}{4}$ \\
\hline
$5$ & $4$ & $ \frac{1}{2}(\theta^{0} + \theta^{3})(1 +
i\theta^{1} \theta^{2}) \theta^5$ & 
$ \frac{3}{4}$ & $ 0$ & $ -i$ & $0$ & $\frac{1}{2} $ 
&$ \frac{3}{4} $ & $-\frac{1}{2}$ & $\frac{3}{4}$ \\
\hline\hline
$6$ & $1$ & $ \frac{1}{2}(1 - \theta^{0}\theta^{3})(\theta^{1} -
i \theta^{2}) \theta^5$ & 
$ \frac{3}{4}$ & $ 1$ & $ 0$ & $0$ & $\frac{1}{2} $ 
&$ \frac{3}{4} $ & $\frac{1}{2}$ & $\frac{3}{4}$ \\
\hline
$6$ & $2$ & $ \frac{1}{2}(\theta^{0} - \theta^{3})(1 +
i\theta^{1} \theta^{2}) \theta^5$ & 
$ \frac{3}{4}$ & $ 0$ & $ i$ & $0$ & $-\frac{1}{2} $ 
&$ \frac{3}{4} $ & $\frac{1}{2}$ & $\frac{3}{4}$ \\
\hline
$6$ & $3$ & $ \frac{1}{2}(1 + \theta^{0} \theta^{3})(\theta^{1}
+ i \theta^{2}) \theta^5$ & 
$ \frac{3}{4}$ & $ -1$ & $ 0$ & $0$ & $-\frac{1}{2} $ 
&$ \frac{3}{4} $ & $-\frac{1}{2}$ & $\frac{3}{4}$ \\
\hline
$6$ & $4$ & $ \frac{1}{2}(\theta^{0} + \theta^{3})(1 -
i\theta^{1} \theta^{2}) \theta^5$ & 
$ \frac{3}{4}$ & $ 0$ & $ -i$ & $0$ & $\frac{1}{2} $ 
&$ \frac{3}{4} $ & $-\frac{1}{2}$ & $\frac{3}{4}$ \\
\hline
\end{tabular}
\end{center}

Table VIb. Irreducible representations of the two subgroups 
$ SU(2) \times SU(2) $ embedded into the group $ SO(1,4) $ as
defined by the generators of vectorial character. There are two
scalars, two three vectors and two four vectors. We present
the eigenvalues of commuting operators $ S^2 = \frac{1}{2} S^{ab}
S_{ab}, S_3 = S^{12}, K_3 = S^{03}, \Gamma, \tau^{13},
(\tau^1)^2, \tau^{23}, (\tau^2)^2.$ 

\vspace{3mm}

\vspace{
3mm}

{\it 4.3.3.  Representations of the group $SU(3)$ embedded in
$SO(6)$ } 

\vspace{3mm}

There are two Casimir operators: $ \it C^1 $  and $
\it C^2 $ (Eq.(3.12)) and two commuting generators: $ \tau^3 $ 
and $ \tau^8 $ (Eq.(3.11)) in this group. We shall be interested
in irreducible representations of vectors with an even Grassmann
character only, taking into account that the part defined by the
group $ SO(1,4) $ determine the Grassmann character of either
fermions or bosons.  

\newpage

\begin{center}
\begin{tabular}{|c|c||c|r|c|}
\hline
$ a$& $i$ & 
$<\theta|\tilde{\varphi}^{a}_{i}>$ & $\tilde{\tau}_{3}$ & 
$\tilde{\tau}_{8}$\\
\hline\hline 
$1$ & $1$ & $ (\frac{1}{2})^{\frac{3}{2}} (1 +
i\theta^{5}\theta^{6}) 
(1-i \theta^{1}\theta^{2})(1 + i\theta^{3} \theta^{4})$
& $\frac{1}{2}$ & $ \frac{1}{2{\sqrt 3}}$\\
\hline
$1$ & $2$ & $ (\frac{1}{2})^{\frac{3}{2}} (1 +
i\theta^{5}\theta^{6}) 
(\theta^{1} + i \theta^{2})(\theta^{3} - i \theta^{4})$
& $- \frac{1}{2}$ & $ \frac{1}{2 {\sqrt 3}}$\\
\hline
$1$ & $3$ & $ -(\frac{1}{2})^{\frac{3}{2}} (\theta^{5} - i
\theta^{6}) 
(\theta^{1} + i \theta^{2})(1 + i \theta^{3}\theta^{4})$
& $0$ & $ -\frac{1}{{\sqrt 3}}$\\
\hline\hline
$2$ & $1$ & $ (\frac{1}{2})^{\frac{3}{2}} (1 +i
\theta^{5}\theta^{6}) 
(\theta^{1} - i \theta^{2})(\theta^{3} + i \theta^{4})$
& $\frac{1}{2}$ & $ \frac{1}{2{\sqrt 3}}$\\
\hline
$2$ & $2$ & $ (\frac{1}{2})^{\frac{3}{2}} (1 + i
\theta^{5}\theta^{6}) 
(1 + i \theta^{1}\theta^{2})(1 - i\theta^{3} \theta^{4})$
& $- \frac{1}{2}$ & $ \frac{1}{2 {\sqrt 3}}$\\
\hline
$2$ & $3$ & $ -(\frac{1}{2})^{\frac{3}{2}} (\theta^{5} - i
\theta^{6}) 
(1 + i\theta^{1} \theta^{2})(\theta^{3}+ i\theta^{4})$
& $0$ & $ -\frac{1}{{\sqrt 3}}$\\
\hline
\hline
$3$ & $1$ & $ (\frac{1}{2})^{\frac{3}{2}} (\theta^{5} +
i\theta^{6}) 
(1-i \theta^{1}\theta^{2})(\theta^{3} + i \theta^{4})$
& $\frac{1}{2}$ & $ \frac{1}{2{\sqrt 3}}$\\
\hline
$3$ & $2$ & $ (\frac{1}{2})^{\frac{3}{2}} (\theta^{5}+ i
\theta^{6}) 
(\theta^{1} + i\theta^{2})(1 -i\theta^{3} \theta^{4})$
& $- \frac{1}{2}$ & $ \frac{1}{2 {\sqrt 3}}$\\
\hline
$3$ & $3$ & $ -(\frac{1}{2})^{\frac{3}{2}} (1 -i
\theta^{5}\theta^{6}) 
(\theta^{1} + i \theta^{2})(\theta^{3} + i\theta^{4})$
& $0$ & $ -\frac{1}{{\sqrt 3}}$\\
\hline
\hline
$4$ & $1$ & $ (\frac{1}{2})^{\frac{3}{2}} (\theta^{5} +
i\theta^{6}) 
(\theta^{1} - i \theta^{2})(1 + i\theta^{3}\theta^{4})$
& $\frac{1}{2}$ & $ \frac{1}{2{\sqrt 3}}$\\
\hline
$4$ & $2$ & $ (\frac{1}{2})^{\frac{3}{2}} (\theta^{5}+ i
\theta^{6}) 
(1 + i \theta^{1} \theta^{2})(\theta^3 -i \theta^{4})$
& $- \frac{1}{2}$ & $ \frac{1}{2 {\sqrt 3}}$\\
\hline
$4$ & $3$ & $ -(\frac{1}{2})^{\frac{3}{2}} (1 -i
\theta^{5}\theta^{6}) 
(1 + i \theta^{1} \theta^{2})(1 + i \theta^{3} \theta^{4})$
& $0$ & $ -\frac{1}{{\sqrt 3}}$\\
\hline
\hline
$5$ & $1$ & $ (\frac{1}{2})^{\frac{3}{2}} (1 -i
\theta^{5}\theta^{6}) 
(1 + i\theta^{1} \theta^{2})(1 -i \theta^{3}\theta^{4})$
& $-\frac{1}{2}$ & $ -\frac{1}{2{\sqrt 3}}$\\
\hline
$5$ & $2$ & $ (\frac{1}{2})^{\frac{3}{2}} (1 -i
\theta^{5}\theta^{6}) 
(\theta^{1} - i \theta^{2})(\theta^{3} + i \theta^{4})$
& $ \frac{1}{2}$ & $ -\frac{1}{2 {\sqrt 3}}$\\
\hline
$5$ & $3$ & $ -(\frac{1}{2})^{\frac{3}{2}} (\theta^{5} +i
\theta^{6}) 
(\theta^{1} - i \theta^{2})(1 -i \theta^{3}\theta^{4})$
& $0$ & $ \frac{1}{{\sqrt 3}}$\\
\hline
\hline
$6$ & $1$ & $ (\frac{1}{2})^{\frac{3}{2}} (1 -i
\theta^{5}\theta^{6}) 
(\theta^{1} + i \theta^{2})(\theta^{3} - i \theta^{4})$
& $-\frac{1}{2}$ & $ -\frac{1}{2{\sqrt 3}}$\\
\hline
$6$ & $2$ & $ (\frac{1}{2})^{\frac{3}{2}} (1 -i
\theta^{5}\theta^{6}) 
(1 -i \theta^{1}\theta^{2})(1 +i \theta^{3}\theta^{4})$
& $ \frac{1}{2}$ & $ -\frac{1}{2 {\sqrt 3}}$\\
\hline
$6$ & $3$ & $ -(\frac{1}{2})^{\frac{3}{2}} (\theta^{5} +i
\theta^{6}) 
(1 -i \theta^{1} \theta^{2})( \theta^{3} - i\theta^{4})$
& $0$ & $ \frac{1}{{\sqrt 3}}$\\
\hline
\hline
$7$ & $1$ & $ (\frac{1}{2})^{\frac{3}{2}} (\theta^{5}- i
\theta^{6}) 
(1 + i \theta^1 \theta^{2})(\theta^{3}- i \theta^{4})$
& $-\frac{1}{2}$ & $ -\frac{1}{2{\sqrt 3}}$\\
\hline
$7$ & $2$ & $ (\frac{1}{2})^{\frac{3}{2}} (\theta^{5}-
i\theta^{6}) 
(\theta^{1} - i \theta^{2})(1 + i\theta^{3} \theta^{4})$
& $ \frac{1}{2}$ & $ -\frac{1}{2 {\sqrt 3}}$\\
\hline
$7$ & $3$ & $ -(\frac{1}{2})^{\frac{3}{2}} (1 + i \theta^{5}
\theta^{6}) 
(\theta^{1} -i \theta^{2})(\theta^{3} - i \theta^{4})$
& $0$ & $ \frac{1}{{\sqrt 3}}$\\
\hline
\hline
$8$ & $1$ & $ (\frac{1}{2})^{\frac{3}{2}} (\theta^{5}- i
\theta^{6}) 
(\theta^{1} + i \theta^{2})(1 -i \theta^{3}\theta^{4})$
& $-\frac{1}{2}$ & $ -\frac{1}{2{\sqrt 3}}$\\
\hline
$8$ & $2$ & $ (\frac{1}{2})^{\frac{3}{2}} (\theta^{5}-
i\theta^{6}) 
(1 -i \theta^{1}\theta^{2})(\theta^{3} + i \theta^{4})$
& $ \frac{1}{2}$ & $ -\frac{1}{2 {\sqrt 3}}$\\
\hline
$8$ & $3$ & $ -(\frac{1}{2})^{\frac{3}{2}} (1 +i \theta^{5}
\theta^{6}) 
(1 -i \theta^{1} \theta^{2})(1 - i \theta^{3}\theta^{4})$
& $0$ & $ \frac{1}{{\sqrt 3}}$\\
\hline
\hline
$9$ & $1$ & $ (\frac{1}{2})^{\frac{3}{2}} (1 -i
\theta^{5}\theta^{6}) 
(1 -i \theta^{1} \theta^{2})(1 -i \theta^{3}\theta^{4})$
& $0$ & $0 $\\
\hline
$10$ & $1$ & $ (\frac{1}{2})^{\frac{3}{2}} (\theta^{5} - i
\theta^{6}) 
(\theta^{1} - i \theta^{2})(1 -i \theta^{3}\theta^{4})$
& $0$ & $0 $\\
\hline
$11$ & $1$ & $ (\frac{1}{2})^{\frac{3}{2}} (\theta^{5} - i
\theta^{6}) 
(1- i \theta^{1} \theta^{2})(\theta^{3} - i\theta^{4})$
& $0$ & $0 $\\
\hline
$12$ & $1$ & $ (\frac{1}{2})^{\frac{3}{2}} (1 -i
\theta^{5}\theta^{6}) 
(\theta^{1} - i \theta^{2})(\theta^{3} - i \theta^{4})$
& $0$ & $0 $\\
\hline
$13$ & $1$ & $ (\frac{1}{2})^{\frac{3}{2}} (1 + i
\theta^{5}\theta^{6}) 
(1 +i \theta^{1} \theta^{2})(1 +i \theta^{3}\theta^{4})$
& $0$ & $0 $\\
\hline
$14$ & $1$ & $ (\frac{1}{2})^{\frac{3}{2}} (\theta^{5} + i
\theta^{6}) 
(\theta^{1} + i \theta^{2})(1 + i \theta^{3} \theta^{4})$
& $0$ & $0 $\\
\hline
$15$ & $1$ & $ (\frac{1}{2})^{\frac{3}{2}} (\theta^{5} + i
\theta^{6}) 
(1 + i \theta^{1}\theta^{2})(\theta^{3} + i \theta^{4})$
& $0$ & $0 $\\
\hline
$16$ & $1$ & $ (\frac{1}{2})^{\frac{3}{2}} (1 + i
\theta^{5}\theta^{6}) 
(\theta^{1} + i \theta^{2})(\theta^{3} + i \theta^{4})$
& $0$ & $0 $\\
\hline

\end{tabular}
\end{center}

Table VIIa: The eight triplets and the eight singlets, the
Grassmann even irreducible representations of the generators of
spinorial character closing the algebra of 
$SU(3)$.  The generators are expressed by the
generators of the Lorentz 
transformations in  $ 6 $ dimensional Grassmann space
(Eq.(3.11)), forming the group $ SO(6)$. Each
triplet is arranged in such a way that the matrix
representations of 
the generators of the group $ 
SU(3)$ coincide with the usual matrix representations of
triplets \cite{grein}. The diagonal matrix elements of
$\tilde{\tau}^3$ and $\tilde{\tau}^8$ are also given.

\vspace{3mm}

Operators of spinorial character define in the space of $2^5$
vectors of an even Grassmann character {\it eight  triplets} and
{\it eight singlets}. We present them in Table VIIa.  
The triplets are arranged in such a way that the matrix
representation of the generators of the group (Eq.(3.11)) agrees
for each triplet with the usual matrix representation
\cite{grein} of the group $ SU(3)$ for either triplets ( the
first 
four triplets ) or for anti triplets ( the second four triplets
). We present in Table VIIa only 
eigenvalues of operators $\tilde{\tau}_3$ and $\tilde{\tau}_8$.

\begin{center}
\begin{tabular}{|c|c||c|r|c|}
\hline
$ a$& $i$ & 
$<\theta|{\varphi}^{a}_{i}>$ & ${\tau}_{3}$ & 
${\tau}_{8}$\\
\hline\hline
$1$ & $1$ & $ (\frac{1}{2})^{\frac{3}{2}} (\theta^{1} +
i\theta^{2}) 
(\theta^{3} - i\theta^{4})(1 + \theta^{5} \theta^{6})$
& $-1$ & $ 0$\\
\hline
$1$ & $2$ & $-(\frac{1}{2})^{\frac{3}{2}} (\theta^{1} -
i\theta^{2}) 
(\theta^{3} + i\theta^{4})(1 + \theta^{5} \theta^{6})$
& $1$ & $ 0$\\
\hline
$1$ & $3$ & $ -\frac{1}{2} (\theta^{1}\theta^{2}- 
\theta^{3} \theta^{4})(1 + \theta^{5} \theta^{6})$
& $0$ & $ 0$\\
\hline
$1$ & $4$ & $-(\frac{1}{2})^{\frac{3}{2}} (\theta^{1} +
i\theta^{2}) 
(1 + \theta^{3} \theta^{4})( \theta^{5} - i \theta^{6})$
& $-\frac{1}{2}$ & $ -\frac{{\sqrt 3}}{2}$\\
\hline
$1$ & $5$ & $-(\frac{1}{2})^{\frac{3}{2}} (\theta^{1} -
i\theta^{2}) 
(1 + \theta^{3} \theta^{4})( \theta^{5} +i \theta^{6})$
& $\frac{1}{2}$ & $ \frac{{\sqrt 3}}{2}$\\
\hline
$1$ & $6$ & $(\frac{1}{2})^{\frac{3}{2}} (1 +
\theta^{1}\theta^{2}) 
(\theta^{3} + i \theta^{4})( \theta^{5} - i \theta^{6})$
& $\frac{1}{2}$ & $ -\frac{{\sqrt 3}}{2}$\\
\hline
$1$ & $7$ & $-(\frac{1}{2})^{\frac{3}{2}} (1 +
\theta^{1}\theta^{2}) 
(\theta^{3} - i \theta^{4})( \theta^{5} + i \theta^{6})$
& $-\frac{1}{2}$ & $ \frac{{\sqrt 3}}{2}$\\
\hline
$1$ & $8$ & $ \frac{1}{2} (1 + \theta^{1}\theta^{2}) 
(\theta^{5} \theta^{6} - \theta^{3} \theta^{4})$
& $0$ & $0$\\
\hline\hline
$2$ & $1$ & $ (\frac{1}{2})^{\frac{3}{2}} (\theta^{1} +
i\theta^{2}) 
(\theta^{3} - i\theta^{4})(1  - \theta^{5} \theta^{6})$
& $-1$ & $ 0$\\
\hline
$2$ & $2$ & $-(\frac{1}{2})^{\frac{3}{2}} (\theta^{1} -
i\theta^{2}) 
(\theta^{3} + i\theta^{4})(1 - \theta^{5} \theta^{6})$
& $1$ & $ 0$\\
\hline
$2$ & $3$ & $-\frac{1}{2} (\theta^{1}\theta^{2}- 
\theta^{3} \theta^{4})(1 - \theta^{5} \theta^{6})$
& $0$ & $ 0$\\
\hline
$2$ & $4$ & $-(\frac{1}{2})^{\frac{3}{2}} (\theta^{1} +
i\theta^{2}) 
(1 - \theta^{3} \theta^{4})( \theta^{5} - i \theta^{6})$
& $-\frac{1}{2}$ & $ -\frac{{\sqrt 3}}{2}$\\
\hline
$2$ & $5$ & $(-\frac{1}{2})^{\frac{3}{2}} (\theta^{1} -
i\theta^{2}) 
(1 - \theta^{3} \theta^{4})( \theta^{5} +i \theta^{6})$
& $\frac{1}{2}$ & $ \frac{{\sqrt 3}}{2}$\\
\hline
$2$ & $6$ & $(\frac{1}{2})^{\frac{3}{2}} (1 -
\theta^{1}\theta^{2}) 
(\theta^{3} + i \theta^{4})( \theta^{5} - i \theta^{6})$
& $\frac{1}{2}$ & $ -\frac{{\sqrt 3}}{2}$\\
\hline
$2$ & $7$ & $-(\frac{1}{2})^{\frac{3}{2}} (1 -
\theta^{1}\theta^{2}) 
(\theta^{3} - i \theta^{4})( \theta^{5} + i \theta^{6})$
& $-\frac{1}{2}$ & $ \frac{{\sqrt 3}}{2}$\\
\hline
$2$ & $8$ & $ - \frac{1}{2} (1 -
\theta^{1}\theta^{2}) 
(\theta^{5} \theta^{6} - \theta^{3} \theta^{4})$
& $0$ & $0$\\
\hline\hline
$3$ & $1$ & $\frac{1}{{\sqrt 2}} (1 + \theta^{1}
\theta^{2}\theta^{3}\theta^{4} 
\theta^{5}\theta^{6})$ & $ 0$& $0$\\
\hline
$4$ & $1$ & $\frac{1}{{\sqrt 2}} (1 -\theta^{1}
\theta^{2}\theta^{3}\theta^{4} 
\theta^{5}\theta^{6})$ & $ 0$& $0$\\
\hline
\end{tabular}
\end{center}

Table VIIb. The two singlets and the two octets, forming the
Grassmann even
irreducible representations of the generators of vectorial
character closing the algebra of $ SU(3)$. The
generators are expressed, as in the case of the operators of
spinorial character, in terms of the generators of the Lorentz
transformations in  six dimensional Grassmann space. Eigenvalues 
of the operators $\tau_3 $ and $\tau_8 $ are also presented. 

\vspace{3mm}

Generators of a vectorial character  in the vector space of
an even Grassmann character define {\it two singlets, two
octets} and 
{\it one multiplet of fourteen vectors}. We present the singlets
and 
octets in Table VIIb, together with the eigenvalues of the 
operators $ \tau^3 $ and $ \tau^8 $, which are diagonal in these
representations. We arrange each of the two octets in such a way
that the eight generators $\tau^i,\;\;i = \{1,8\} $ from
Eq.(3.11), 
when $ M^{ab} $ are replaced by $ S^{ab}$, have equal matrix
representations for both octets. These matrix representations
are in agreement with  the matrices defined by
the structure constants of the group $ SU(3)$ $ (T^i)_{jk} = - i
f^{1ijk} $, $ f^{1ijk}$ taken from Table I, if
transformed into the basis in which $ T^3 $ and $ T^8 $ are
diagonal. 

While triplets and singlets, determined by the generators of
spinorial 
character of the group $ SU(3) $, can be either identified with
fundamental representations of the group (triplets), or  can be
used to describe fermions
without the colour charge (singlets), can octets and singlets,
determined by the generators of the vectorial character of the
same group, be identified with adjoint representations of the
group (octets)
or can be used to describe bosons without charges (singlets).

\vspace{3mm}

{\it 4.3.4. Representations of the group SO(1,7) in terms of
subgroups $ SO(1,4) \times SU(2)$ }

\vspace{3mm}

The representations of the group $ SO(1,4) \times SU(2) $ ,
embedded into the Lorentz group $ SO(1,7) $ can be deduced from
the representations presented in the two Subsect. $4.3.1.$ and
$4.3.2.$. In the second subsection we found the irreducible 
representations of the group $ SO(1,4)$, defined in the vector
space spanned over the coordinate Grassmann space $\theta^a,\;\; a
= \{0,1,2,3,5\}$.  The vector space of an
odd Grassmann character was analyzed with respect to generators
of spinorial character. We found eight bispinors arranged into
four four spinors. The vector space of an even Grassmann
character was analyzed with respect to generators of vectorial
character. We found two scalars, two three vectors and two four
vectors. The fifth dimension was needed to define the $\gamma^a$
matrices.

The vector space spanned over the coordinate space of
$\theta^a,\;\; a = \{6,7,8\} $, when analyzed with
respect to operators of spinorial character, gives {\it two
doublets} of an {\it even} Grassmann character ( Subsect.
$4.3.1.$, if the indices 1,2,3 are replaced by the indices 6,7,8
). In the same subspace the operators of vectorial 
character define {\it one scalar} and {\it one three vector} of
an {\it even} Grassmann character.  

In the space of eight Grassmann coordinates there exist, with
respect to the groups $SO(1,4) \times SU(2) $ embeded into the
group $ SO(1,7) $ {\it two
times eight bispinors} ( each bispinor has two vectors ), which
are {\it doublets} with respect 
to the group $ SU(2) $ and have an {\it odd} Grassmann
character, as well as 
{\it two scalars} and {\it two three vectors} of an even Grassmann
character, which are either {\it triplets} or {\it singlets}
with respect to the group $ SU(2)$. Each bispinor ( with respect
to the group $SO(1,3) $ ), which is a doublet ( with respect to
the group $ SU(2) $ ) 
represents the fundamental representation ( with respect to both
subgroups ), while accordingly a three vector ( with respect to
the group $SO(1,3)$ ), which is a triplet ( with respect to the
group $ SU(2) $ ) represents the adjoint 
representation with respect to both subgroups. There exists no
bispinor which would be a singlet with respect to the group 
$ SU(2) $ in this scheme.

\vspace{3mm}

{\it 4.3.5. Representations of $SU(3) \times SU(2) \times U(1) $
embedded into the group $ SO(10) $ }

\vspace{3mm}

We look for the irreducible representations of generators
$\tau^{\alpha i},\;\; \alpha = 1,2,3,\;\;i = 1,n_{\alpha} $,
expressed in terms of the generators of the Lorentz group $
SO(10)$ as defined in Eqs.(3.11) and (3.14). We shall rename
the operators $ \tau_i, i = 1,2,..,8 $ from Eq.(3.11) into $
\tau^{1i} , i = 1,2,..,8 $.

\vspace{3mm}

{\bf $SU(3)$}

\vspace{3mm}

The representations of the
group $SU(3)$ were already presented in Subsect.4.3.3..

\vspace{3mm}

{\bf $SU(2)$}

\vspace{3mm}

The representations of the group $ SU(2)$ were presented in
Subsect. 4.3.1., since the algebras of $SO(3)$ and $SU(2)$ are
isomorphic. However, since the generators of $SU(2)$ are now
written in terms of four generators of the Lorentz group rather
than in terms of three (see Eq.(3.14a)), we have to solve the
eigenvalue problem again.

Solving the eigenvalue problem (Eq.(4.6)) for the operators of
spinorial character $\tilde\tau^{2i},\;\;i = \{1,2,3\} $ (
Eq.(3.14a)), expressed in terms of $ \tilde S^{ab},\;\;a,b
=\{7,8,9,10\} $, we find {\it two doublets} and {\it four
singlets} of an {\it even} Grassmann character. We present these
vectors together with the eigenvalues of the commuting operators in
Table VIIIa. 

`
\begin{center}
\begin{tabular}{|c|c||c|r|c|}
\hline
$ a$& $i$ & 
$<\theta|\tilde{\varphi}^{a}_{i}>$ & $\tilde{\tau}^{23}$ & 
$(\tilde{\tau}^{2})^{2}$\\
\hline\hline 
$1$ & $1$ & $ \frac{1}{2} (1 - i \theta^{7} \theta^{8}) 
(1 + i \theta^{9} \theta^{10})$
& $\frac{1}{2}$ & $ \frac{3}{4}$\\
\hline
$1$ & $2$ & $ \frac{1}{2} (\theta^{7} + i \theta^{8}) 
(\theta^{9} - i \theta^{10})$
& $-\frac{1}{2}$ & $ \frac{3}{4}$\\
\hline\hline
$2$ & $1$ & $ \frac{1}{2} (1 + i \theta^{7} \theta^{8}) 
(1 - i \theta^{9} \theta^{10})$
& $- \frac{1}{2}$ & $ \frac{3}{4}$\\
\hline
$2$ & $2$ & $ \frac{1}{2} (\theta^{7} - i \theta^{8}) 
(\theta^{9} + i \theta^{10})$
& $\frac{1}{2}$ & $ \frac{3}{4}$\\
\hline\hline
$3$ & $1$ & $ \frac{1}{2} (1 + i \theta^{7} \theta^{8}) 
(1 + i \theta^{9} \theta^{10})$
& $0$ & $0$\\
\hline
$4$ & $1$ & $ \frac{1}{2} (\theta^{7} + i \theta^{8}) 
(\theta^{9} + i \theta^{10})$
& $0$ & $ 0$\\
\hline
$5$ & $1$ & $ \frac{1}{2} (1 - i \theta^{7} \theta^{8}) 
(1 - i \theta^{9} \theta^{10})$
& $0$ & $0$\\
\hline
$6$ & $1$ & $ \frac{1}{2} (\theta^{7} - i \theta^{8}) 
(\theta^{9} - i \theta^{10})$
& $0$ & $ 0$\\
\hline
\end{tabular}
\end{center}

Table VIIIa: The irreducible representations of the generators of
spinorial character $ \tilde{\tau} ^{2i},\;\;i = \{1,2,3\} $,
defined in Eqs.(3.14a) with $ M^{ab}$  replaced by $ \tilde
S^{ab} $, closing the algebra of $SU(2)$.
In the vector space of an even Grassmann character there are two
doublets and four singlets. The expectation values of the
commuting 
operators are also added. 

\vspace{3mm}

Generators of vectorial character $\tau^{2i},\;\;i = \{1,2,3\}
$, expressed in terms of $ S^{ab},\;a,b = \{7,8,9,10\}$
(Eq.(3.14a)), 
define {\it one three vector} and {\it five singlets} of an
{\it 
even} Grassmann character. We present the vectors and the
eigenvalues 
of the commuting operators in Table VIIIb.

\begin{center}
\begin{tabular}{|c|c||c|r|c|}
\hline
$ a$& $i$ & 
$<\theta|{\varphi}^{a}_{i}>$ & ${\tau}^{23}$ & 
$({\tau}^{2})^{2}$\\
\hline 
$1$ & $1$ & $ \frac{1}{2} (\theta^{7} + i\theta^{8}) 
(\theta^{9} - i\theta^{10})$
& $-1$ & $ 1$\\
\hline
$1$ & $2$ & $ \frac{1}{2} (\theta^{7} - i \theta^{8}) 
(\theta^{9} + i \theta^{10})$
& $1$ & $ 1$\\
\hline
$1$ & $3$ & $ \frac{-i}{\sqrt{2}} ( \theta^{7} \theta^{8} - \theta^{9}
\theta^{10})$ 
& $0$ & $ 1$\\
\hline\hline
$2$ & $1$ & $ \frac{1}{\sqrt{2}} (1 +
\theta^{7}\theta^{8}\theta^{9} \theta^{10})$ 
& $0$ & $ 0$\\
\hline
$3$ & $1$ & $ \frac{1}{\sqrt{2}} (1 -
\theta^{7}\theta^{8}\theta^{9} \theta^{10})$ 
& $0$ & $0$\\
\hline\hline
$4$ & $1$ & $ \frac{1}{2} (\theta^{7} + i \theta^{8}) 
(\theta^{9} + i \theta^{10})$
& $0$ & $ 0$\\
\hline
$5$ & $1$ & $ \frac{1}{2} (\theta^{7} - i \theta^{8}) 
(\theta^{9} - i \theta^{10})$
& $0$ & $0$\\
\hline
$6$ & $1$ & $ \frac{-i}{\sqrt{2}} (\theta^{7} \theta^{8} +
\theta^{9}\theta^{10})$ 
& $0$ & $ 0$\\
\hline
\end{tabular}
\end{center}

Table VIIIb. The irreducible representations of the generators
of vectorial character $ \tau ^{\alpha i}, \;\;i = \{1,2,3\}$,
defined in Eqs.(3.14a) if $  M^{ab}$ are replaced by $ 
S^{ab} $, and closing the algebra of $SU(2)$.
In the Grassmann even part of the space there are two doublets
and four singlets. The expectation values of the commuting
operators 
are also presented.

\vspace{3mm}

\vspace{3mm}

{\bf $U(1)$}

\vspace{3mm}

Solving the eigenvalue problem for the generator $\tau^{31}$
(Eq.(3.14b)) of the group $U(1)$ either for operators of
spinorial character (when $M^{ab} = \tilde
S^{ab}$) or of vectorial character (when $M^{ab} = S^{ab}$), we
find that the products of vectors from 
Table VIIa ( VIIb) forming the irreducible representations of
$SU(3)$ 
with vectors from Table VIIIa (VIIIb),
forming the irreducible representations of $ SU(2)$, 
for either of the two types of  operators are  the
eigenvectors of the operator $\tau^{31}$ as well. We have, of
course, to choose only one set of vectors   from each
table  and make the outer products among them.
We always combine the  vectors of Table
VIIa with the vectors of Table VIIIa, and the vectors of Table
VIIb with the vectors of Table VIIIb.  

When looking for  representations of the operator $\tilde
\tau^{31}$ (Eq.(3.14b)) of
spinorial character we find that   products of triplets from
Table VIIa with doublets from Table VIIIa have two different
expectation 
values of the operator $\tilde \tau^{31}$ defining the group
$U(1)$. Products of triplets from Table 
VIIa with singlets from Table VIIIa have four
different eigenvalues of $\tilde \tau^{13}$.
Products of singlets from Table VIIa with doublets of Table
VIIIa have two different eigenvalues of
$\tilde \tau^{13}$, while products of singlets of Table 
VIIa with singlets of Table VIIIa have three different
eigenvalues 
of $\tilde \tau^{13}$.
We present these values in Table IXa.

\begin{center}
\begin{tabular}{|c|c|c|}
\hline
$<\theta|\tilde{\varphi}^{a}_{i}>$&
$<\theta|\tilde{\varphi}^{b}_{j}>$& 
$\tilde{\tau}^{31}$\\
\hline 
$a \in \{1,..,8\}$ &  $b \in \{1,2\}$ & $\enspace$\\
$ i\in \{1,2,3\}$ & $ j \in \{1,2\}$ & ${\sqrt \frac{3}{5}}
\frac{1}{6},\enspace a \in \{1,..,4 \}$\\
$\in triplets$ & $\in doublets $ & ${- \sqrt \frac{3}{5}}
\frac{1}{6}, a \in \{5,..,8\}$\\
\hline
$a \in \{1,..,8\}$ & $b \in \{3,..,6\}$ & $-{\sqrt
\frac{3}{5}} \frac{1}{3},\enspace a \in \{1,..,4 \}, \enspace b
\in \{3,4\}$\\
$ i\in \{1,2,3\}$ & $ j \in \{1\}$ & ${\sqrt \frac{3}{5}}
\frac{2}{3}, \enspace a \in \{1,..,4\}, \enspace b \in
\{5,6\}$\\ 
$\in triplets$ & $\in singlets $ & ${-\sqrt \frac{3}{5}}
\frac{2}{3}, \enspace a \in \{5,..,8\}, \enspace b \in
\{3,4\}$\\ 
$\enspace$ & $\enspace $ & ${\sqrt \frac{3}{5}} \frac{1}{3},
\enspace a \in \{5,..,8\}, \enspace b \in \{5,6\}$\\
\hline
$a \in \{9,..,16\}$ &  $b \in \{1,2\}$ & $\enspace$\\
$ i \in \{1\}$ & $ j \in \{1,2\}$ & $-{\sqrt \frac{3}{5}}
\frac{1}{2},\enspace a \in \{9,..,12 \}$\\ 
$\in singlets$ & $\in doublets $ & ${\sqrt \frac{3}{5}}
\frac{1}{2},\enspace a \in \{13,..,16 \}$\\ 
\hline
$a \in \{9,..,16\}$ & $b \in \{3,..,6\}$ & $-{\sqrt
\frac{3}{5}},\enspace a \in \{9,..,12 \}, 
\enspace b \in \{3,4\}$\\
$ i \in \{1\}$ & $ j \in \{1\}$ & 
$0, \enspace a \in \{9,..,12 \}, \enspace b \in \{5,6 \}$\\
$\in singlets$ & $\in singlets $ & $0, \enspace a \in
\{13,..,16\}, \enspace b \in \{3,4\}$\\ 
$\enspace$ & $\enspace$ & ${\sqrt \frac{3}{5}},\enspace a \in
\{13,..,16\}, \enspace b \in \{5,6\}$\\ 
\hline
\end{tabular}

\end{center}

Table IXa. Eigenvalues of the operator $\tilde \tau ^{31}$
forming 
the group of $U(1)$ (Eq.(3.14b)), with $M^{ab} = \tilde S^{ab}
$. 
Eigenvectors of $\tilde \tau 
^{31}$ are the outer products of vectors from Table VIIa and
Table VIIIa. The first column concerns vectors from Table VIIa,
which form either triplets or singlets. The second column
concerns vectors from Table VIIIa. They form either doublets or
singlets. 

\vspace{3mm}

When looking for the representations of the operator $\tau^{31}$
(Eq.(3.14b)) of vectorial character, we find that products of
octets from Table VIIb with triplets from Table VIIIb and products
of singlets from Table VIIb with triplets from Table VIIIb have
expectation values of the operator $\tau^{31}$ equal to zero.
Products of octets from Table VIIb with singlets from Table VIIIb
and products of singlets from Table VIIb with singlets from Table
VIIb have expectation values of $\tau^{31}$ zero or $\pm 1$. We
present these values in Table IXb.

\begin{center}
\begin{tabular}{|c|c|l|}
\hline
$<\theta|{\varphi}^{a}_{i}>$& $<\theta|{\varphi}^{b}_{j}>$& 
${\tau}^{31}$\\
\hline 
$a \in \{1,2\}$ &  $b \in \{1\}$ & $\enspace$\\
$ i \in \{1,...,8\}$ & $ j \in \{1,2,3\}$ & $0$\\
$\in octets$ & $\in triplets $ & $\enspace$\\
\hline
$a \in \{1,2\}$ & $b \in \{2,...,6\}$ & $0, \enspace b
\in\{2,3,6\},\enspace 
j \in \{1\}$\\
$i \in \{1,...,8\}$ & $ j \in \{1\}$ & $\sqrt{\frac{3}{5}},
\enspace b 
\in\{4\},\enspace 
j \in \{1\}$\\
$\in octets$ & $\in singlets $ & $-\sqrt{\frac{3}{5}}, \enspace
b \in 
\{5\},\enspace 
j \in \{1\}$\\
\hline
$a \in \{3,4\}$ &  $b \in \{1\}$ & $\enspace$\\
$ i \in \{1\}$ & $ j \in \{1,2,3\}$ & $0$\\
$\in singlets$ & $\in triplets $ & $\enspace$\\
\hline
$a \in \{3,4\}$ & $b \in \{2,...,6\}$ & $0, \enspace b
\in\{2,3,6\},\enspace 
j \in \{1\}$\\
$ i\in \{1\}$ & $ j \in \{1\}$ & $\sqrt{\frac{3}{5}}, \enspace b
\in\{4\},\enspace 
j \in \{1\} $\\
$\in singlets$ & $\in singlets $ & $ -\sqrt{\frac{3}{5}},
\enspace b 
\in\{5\},\enspace 
j \in \{1\}$\\
\hline
\end{tabular}
\end{center}

Table IXb. Eigenvalues of the operator $\tau^{31}$ from
Eq.(3.14b), 
with $M^{ab} = S^{ab} $. The operator $\tau^{31}$ has a
vectorial character. We choose 
representations which are the outer product of vectors from
Table 
VIIb with vectors of Table VIIIb.  The first column concerns
vectors 
from Table VIIb, 
which form either octets or singlets. The second column
concerns vectors from Table VIIIb. They form either triplets or 
singlets.

\vspace{3mm}

\vspace{3mm}

{\it 4.3.6. Represenations of the group $SO(1,14)$ in terms of
$SU(3) 
\times SU(2) \times U(1) $ }

\vspace{3mm}

We find the representations of the group $SO(1,14)$ as the outer
products of the representations of the subgroup $SO(1,4)$ 
and the subgroups $SU(3), SU(2)$ and $U(1) $. The representations
of these groups are discussed in
Subsect. 4.3.2. and 4.3.5..  We look for the spinorial and 
the vectorial representations in $15 $ dimensional Grassmann space.  
The former are defined by the spinorial operators $\tilde S^{ab}$, 
the latter by the  vectorial operators $ S^{ab}$. In the
spinorial case we choose vectors of an odd Grassmann
character, while  in the vectorial case we choose vectors of an
even Grassmann character. The reason for that is the requirement
that in the canonical quantization of fields the former quantize
to fermions, the later to bosons \cite{man2}. Since
the character of fields, that is their behaviour as spinors,
vectors or scalars, is determined by the behaviour of fields
with respect to generators of the Lorentz transformations in
the four dimensional subspace of the fifteen dimensional space,
we choose vectors  to be either of an odd or of an even
Grassmann character with respect to the group $SO(1,4)$, where
the fifth dimension is needed to properly define the
Grassmann character of the fields as well as the Dirac $\tilde
\gamma ^a$ matrices.
That part of the Grassmann space, which determines charges of
fields,  is connected with the group $SO(10)$. It is 
spanned over  coordinate space with indices higher then five.
It is chosen to participate to the outer products of vectors by
the part 
of an even Grassmann character only, either for the spinorial or
for the vectorial case. The part of an odd Grassmann character
will be studied elswhere\cite{ana}. It turns out, for example,
that in an outer product with representations of the group
$SO(1,4)$ of an odd Grassmann character, this part offers in the
vectorial case three vectors, which are doublets with respect to
the group $ SU(2)$.

Representations of the group $SO(1,14)$ are  the outer
products of the representations of the group $SO(1,4)$ and of the
group $SO(10)$ discussed in the previous subsections. They all
belong either to the spinorial representations or to the
vectorial representations. There are spinorial representations
which coincide with what is called the fundamental
representations with respect to the group $ SO(1,3) $ and the
groups $ SU(3), SU(2)$ and $ U(1) $, but there are also
representations which are singlets with respect to any of groups
$SU(3)$ or $SU(2). $
And there are the vectorial representations which coincide with the
adjoint representations with respect to all or some of the groups
$SO(1,3), SU(3), SU(2) $, some of them are singlets with respect
to some or all these groups.

{\it The structure of Grassmann space, with the limited number of
vectors limits the possible 
representations allowed by the group theory.}

\vspace{4mm}

{\it 5. Concluding remarks}

\vspace{3mm}

In this paper the  algebras and
subalgebras defined by  two kinds of  generators  of the
Lorentz transformations, forming in $15 $ dimensional Grassmann
space the group $ SO(1,14) $,
one of spinorial, the other of vectorial character, both  the
linear differentional operators, 
were studied and some of their representations were obtained,
those, which have in $ SO(1,4) $ an odd ( when describing
fermions) or an even ( when describing bosons) Grassmann
character, while in $ SO(10) $ they have only an even Grassmann
character. 
According to two kinds of generators 
defined in the linear vector space, spanned over  Grassmann
coordinate space, there are also two kinds of representations: 
we call them spinorial and vectorial representations,
respectively. We find among spinorial representations the 
representations, which are known as the fundamental
representations, and we find also singlets,
needed to describe fermions without a particular
charge. Among vectorial representations we find accordingly
the adjoint or the regular
representations, and again we find also singlets, which are
needed to describe bosons without a particular charge.

The  
generators of translations of an odd Grassmann character were
used  to find the decomposition of the Lorentz group 
$SO(2n)$ in terms of the subgroup $SU(n)$. This decomposition
can be found either for operators of spinorial
or for operators of vectorial character. 

We looked for irreducible representations of the 
operators closing subalgebras and of the corresponding Casimir 
operators to find for each of subalgebras two kinds of
representations, the spinorial and the vectorial onces in
accordance with the two
kinds of operators.
Since the dimension of the vector space, spanned over 
Grassmann coordinate space, is finite, all representations are
finite dimensional.

Since the group $ SO(1,d-1) $ contains for $ d=15 $ as subgroups
the groups $ 
SO(1,3)$, needed to describe spins of fermions and bosons, as
well as $U(1), SU(2) $ and $ SU(3) $, needed to describe the
Yang-Mills charges of fermions and bosons, {\it the spin and the
Yang - Mills charges of either fermions or of bosons are in the
presented approach unified}. Since spins and charges are described
by the representations of the generators of the Lorentz
transformations of either fermionic or of bosonic character, it
means that {\it fermionic  states must belong either to the
fundamental representations with respect to the groups,
describing charges, or they must be
spinorial singlets  with respect to those groups, while
bosonic states must belong either to the adjoint representations
with respect to the groups, describing charges, or they must be
vectorial singlets with respect to those groups}. One can
find\cite{ana} for vectorial case besides octets and singlets of
$SU(3)$ also 
triplets and besides triplets and singlets of $SU(2)$ also
doublets. But these representations have an odd Grassmann
character in $ SU(10)$ and are not studied in this
article\cite{ana}.  

Among  representations of the proposed approach are 
 the ones, needed to 
describe the quarks, the leptons and
the gauge bosons, which appear in the Standard Electroweak
Model. 
We find left handed spinors, $SU(3)$ triplets and $SU(2)$
doublets with $U(1)$ charge  $\pm \frac{1}{6}$ and
right handed spinors, $SU(3)$ triplets and $SU(2)$ singlets 
with $U(1)$ charge $\pm \frac{2}{3}$ and $ \mp \frac{1}{3}$, which
describe quarks. We find left handed spinors, $SU(3)$ singlets
and $SU(2)$ doublets with $U(1)$ charge $\mp \frac{1}{2}$
and right handed spinors, $SU(3)$ singlets and $SU(2)$ singlets
with $U(1)$ charge $\mp 1$, which describe leptons.
We find left and right handed three vectors, $SU(3)$ triplets
and $SU(2)$ singlets with $U(1)$ charge $0$, describing gluons
and left handed $SU(3)$ singlets and $SU(2)$ triplets with
$U(1)$ charge $0$, describing massless weak bosons and left
and right handed $SU(3)$ singlets and $SU(2)$ singlets
with $U(1)$ charge $0$, describing  a $U(1)$
field.
 The Higgs's boson of this model
appears due to the presented representations as a
constituent field, while within above mentioned odd
representations 
it appears\cite{ana}
as a scalar, which is  a $SU(3)$ singlet and
$SU(2)$ doublet,  with an odd
Grassmann character in $SO(1,4) $ and $SU(2) $ part of the
Grassmann space.

 The
supersymmetic partners of  the gauge
bosons, required by the supersymmetric extension of the
Standard Electroweak Model, can in the proposed theory exist only
as constituent particles.

In ref. \cite{man1,man2} we presented the action from which it
follows that the unification of spins and charges in Grassmann
space leads to the unification of all interactions: the
gravitational fields in the space of d ordinary and d Grassmann
coordinates may manifest under certain conditions
in the four dimensional subspace of the d dimensional
space as the ordinary gravity and the gauge fields, both
depending on ordinary and Grassmann coordinates, if $ d\ge 15 $.
Grassmann 
coordinates describe all the internal degrees of freedom of
fields - 
their spins and charges. 
The generators of the Lorentz transformations in Grasssmann
space, defining the Yang - Mills charges, commute with the
generators of the Lorentz transformations in the four
dimensional subspace in accordance with the Coleman - Mandula
theorem \cite{col} as well as with it extension for the
supersymmetric case \cite{haag}.

The proposed approach also offers
representations, not included in the Standard
Electroweak Model. These representations, which predict
physics beyond the Standard Electroweak Model, need a further
study\cite{ana}.

\vspace{3mm}

{\bf 5.Acknowledgement. } The work was supported by Ministry of 
Science and Technology of Slovenia. One of the authors (
N.M.B.)would like to 
thank C. Froggatt, H.B. Nielsen and M. Polj\v sak for fruitful
discussions, 
A.Bor\v stnik for checking carefully tables of representations
and G. Veneziano for kind hospitality in CERN, where this paper
was finished.

\end{document}